\documentclass[10pt]{article}

\usepackage{arxiv}
\usepackage[utf8]{inputenc} 
\usepackage[T1]{fontenc}    
\usepackage{hyperref}       
\usepackage{url}            
\usepackage{booktabs}       
\usepackage{amsfonts}       
\usepackage{amssymb}       
\usepackage{nicefrac}       
\usepackage{microtype}      
\usepackage{lipsum}
\usepackage{fancyhdr}       
\usepackage{graphicx}       
\usepackage{titling}       
\usepackage[version=4]{mhchem}
\usepackage{stmaryrd}
\usepackage{microtype}              
\graphicspath{{media/}}     

\usepackage{listings}
\usepackage{xcolor}

\hypersetup{
    colorlinks=true,
    linkcolor=blue,
    filecolor=magenta,      
    urlcolor=cyan,
    pdfauthor={Jay Kuri},
    pdftitle={Vouchsafe: A Zero-Infrastructure Capability Graph Model for Offline Identity and Trust},
    pdfsubject={Vouchsafe: ZI-CG},
    pdfstartview={FitH} 
}
\title{Vouchsafe: A Zero-Infrastructure Capability Graph Model for Offline Identity and Trust}
\author{
    Jay Kuri\\
    Ionzero Inc.\\
    \href{mailto:jayk@ionzero.com}{jayk@ionzero.com}
}
\date{December 22nd, 2025}

\begin{document}
\maketitle

\begin{abstract}

Modern identity and trust systems collapse in the environments where they are
needed most: disaster zones, disconnected or damaged networks, and adversarial
conditions such as censorship or infrastructure interference. These systems
depend on functioning networks to reach online authorities, resolvers,
directories, and revocation services, leaving trust unverifiable whenever
communication is unavailable or untrusted. This work demonstrates that secure
identity and trust are possible without such infrastructure. We introduce the
Zero-Infrastructure Capability Graph (ZI-CG), a model showing that identity,
delegation, and revocation can be represented as self-contained, signed
statements whose validity is determined entirely by local, deterministic
evaluation. We further present Vouchsafe, a complete working instantiation of
this model built using widely deployed primitives including Ed25519, SHA-256,
and structured JSON Web Tokens, requiring no new cryptography or online
services. The results show that a practical, offline-verifiable trust substrate
can be constructed today using only the cryptographic data presented at
evaluation time.
\end{abstract}

\section{Introduction}

\subsection{Motivation}
Identity and trust underpin every security decision in distributed systems.
Whether accepting a message, granting access, or delegating authority, a system
must reliably determine \textbf{who} an actor is and \textbf{whether they are
authorized to perform a given action}. Today, these determinations almost
universally depend on external infrastructure: certificate authorities, DNS,
identity providers, directory services, DID resolvers, revocation endpoints,
and other online sources of truth. In this paradigm, an identity is not
self-contained; it is a reference that must be resolved elsewhere.

However, the environments in which secure identity and trust matter most, field
operations, disaster response, ad hoc peer networks, intermittently connected
environments, and adversarial conditions, are precisely those in which
infrastructure cannot be assumed. Networks may be offline, overloaded,
partitioned, compromised, or intentionally obstructed. In these cases, identity
cannot be looked up, and trust information cannot be retrieved, leaving
traditional systems unable to function.

\subsection{The Core Problem: Infrastructure as Attack Surface}
Beyond fragility, these dependencies introduce a deeper structural
vulnerability: \textbf{every piece of infrastructure in the identity and trust supply
chain is itself an attack surface.} If any component that a verifier must
consult can be intercepted, blocked, spoofed, poisoned, coerced or otherwise
interfered with then the correctness of identity and trust evaluation becomes
uncertain. Historical compromises of certificate authorities, failures of
revocation infrastructure, keyserver poisoning, resolver hijacking, and
widespread censorship illustrate that the infrastructure required to validate
trust is routinely the weakest element of the trust model.

The problem is therefore not only availability but \textbf{epistemology}: the
correctness of an identity claim becomes conditioned on the correctness and
reachability of the supporting infrastructure. As long as verification depends
on external resolution, identity and trust cannot be guaranteed in the presence
of outages, misconfigurations, or adversarial interference.

\subsection{Key Insight: Trust Should Be Verifiable From the Data Alone}
These limitations motivate a different approach: \textbf{identity and trust should be
verifiable directly from the data presented}, without reliance on 
infrastructure such as registries, resolvers, directories, online status
information, or any form of synchronized global state. For identity and trust
to remain reliable across network partitions, organizational boundaries, and
adversarial environments, they must be:

\begin{itemize}
  \item \textbf{self-initiated} - no authority must grant or register an identity
  \item \textbf{self-contained} - verification must require no online lookup; and
  \item \textbf{locally decidable} - any party must be able to evaluate trust using only the presented data.
\end{itemize}

This observation leads to the concept of a \textbf{Zero-Infrastructure Capability
Graph (ZI-CG)}: a class of trust models in which identity is self-verifying
and all trust relationships: delegation, attestation, revocation, and identity
termination, are expressed as signed, content-addressed statements forming a
deterministic directed acyclic graph. Trust evaluation becomes a pure offline
computation over this graph rather than a process mediated by infrastructure.

A central distinction from traditional identity systems is that ZI-CG treats
the signed statement as the fundamental unit of trust. In concrete systems
these signed statements take the form of tokens that can be validated without
external infrastructure. Traditional identity systems assign trust directly to
principals, but a ZI-CG system treats each token as a concrete, verifiable act
of assertion or delegation. Trust is therefore evaluated over the statements
that identities make, not over the identities in isolation.

Importantly, this does not eliminate identity; instead, ZI-CG defines identity
as a cryptographically self-verifying actor capable of making authenticated
statements about itself and others, with trust derived from explicit
endorsement of those statements rather than from accounts or
infrastructure-maintained records.

In this work, we treat identity and trust as primitives that must remain
correct under the most constrained operational conditions. Verifiers may have
no network access, no contact with trusted infrastructure, no ability to fetch
revocation state, and no assurance that any resolver, registry, or
synchronization mechanism is available or trustworthy. Trust decisions must
therefore be computable using only locally available data.

The ZI-CG model shows that the combined problem of identity, delegation, and
revocation, typically treated as dependent on online infrastructure, admits a
fully deterministic offline formulation. In this framework, all trust decisions
reduce to cryptographic verification and evaluation of a finite,
content-addressed graph of signed statements, with no reliance on external
resolution, mutable global state, or third-party authorities. This establishes
that correctness of identity and trust can be derived directly and exclusively
from the presented data.

In this work, we adopt these constraints as the foundation for a general model
of offline-verifiable identity and trust. We formalize the ZI-CG model and
present \textbf{Vouchsafe} as the first complete implementation of this model,
demonstrating that a fully self-contained trust substrate can be constructed
using standard, widely deployed cryptographic primitives.

\subsection{Contributions}

This paper makes two complementary contributions:

\begin{enumerate}
  \item \textbf{The Zero-Infrastructure Capability Graph (ZI-CG) model.}\\
We formalize a trust and authorization framework in which identity,
attestation, delegation, revocation, and identity termination are represented
as self-contained, hash-linked cryptographic statements. A system qualifies as
ZI-CG compliant when all trust-relevant state is encoded in signed tokens and
trust evaluation is a deterministic local computation over a finite token set,
without reliance on online infrastructure.

  \item \textbf{A complete operational instantiation in Vouchsafe.}\\
To the authors' knowledge, Vouchsafe is the first system to fully implement the
ZI-CG model. Vouchsafe provides self-verifying identities, scoped attestation
and delegation tokens, explicit revocation and identity burn semantics, and a
deterministic offline evaluation procedure based solely on standard
cryptographic primitives.
\end{enumerate}

In support of these contributions, this work:

\begin{itemize}
  \item \textbf{Defines a self-verifying identity construction} that enables
identity validation without lookup services, resolver infrastructure, or
preexisting directories.

  \item \textbf{Introduces a self-authenticating token model} in which issuer
identity, statement authenticity, and content integrity are jointly
verifiable using only the token contents.

  \item \textbf{Formalizes a capability-based trust graph} composed of
attestation and delegation statements, with explicit revocation and identity
termination semantics enforced through data commitment rather than mutable
state.

  \item \textbf{Establishes acyclic delegation by construction}, deriving a
strict partial order over statements from content-addressed references, and
thereby eliminating delegation cycles without requiring external detection or
global coordination.

  \item \textbf{Specifies a deterministic authorization semantics} based on
monotonic scope attenuation and minimum-required capability satisfaction,
where authority is evaluated along individual delegation paths and never
expanded through path combination.

  \item \textbf{Provides a fully offline evaluation procedure} that performs
state resolution, capability graph construction, and authorization decisions
using only local inputs, independent of search strategy or evaluation order.

  \item \textbf{Analyzes the security properties} of both the model and its
implementation, including resistance to substitution attacks, graph
manipulation, replay, and infrastructure compromise.
\end{itemize}

Together, these contributions demonstrate that a fully self-contained,
zero-infrastructure trust substrate is not only conceptually coherent but
practically realizable. The following section situates this approach relative
to prior work, highlighting differences in assumptions about identity
resolution, infrastructure dependence, and trust composition.

\section{Background and Related Work}

Decentralized trust, identity binding, and distributed authorization appear in
a wide range of systems with differing assumptions about infrastructure,
identity resolution, and delegation semantics. The Zero-Infrastructure
Capability Graph (ZI-CG) model builds on ideas present in multiple strands of
work, including SPKI, Web-of-Trust systems, key-centered identity
representations, and capability-based authorization. This section summarizes
the major conceptual frameworks that form the landscape in which ZI-CG is
situated.

\subsection{Identity and Trust Models}

\subsubsection*{Public Key Infrastructure (PKI)}
Traditional PKI systems bind identities to public keys using X.509 certificates
issued by hierarchical certificate authorities. Verification relies on
evaluating certificate chains and on infrastructure for assessing revocation
status, such as Certificate Revocation Lists (CRLs) and Online Certificate
Status Protocol (OCSP) services, as profiled in RFC 5280. \cite{Cooper2008PKI}
PKI provides a global naming and trust hierarchy, and identity is defined
externally. PKI illustrates how institutional trust anchors can structure
identity binding and global interoperability.  

\subsubsection*{PGP and the Web-of-Trust}
PGP systems build trust through user-to-user signatures, forming a
decentralized "Web-of-Trust" in which individuals sign each other's keys.
Identity representations typically combine keys with human-readable identifiers
such as email addresses and names, and key discovery and revocation commonly
rely on public keyservers. \cite{ULRICH2011} PGP demonstrates how decentralized trust
relationships can be constructed socially, without centralized authorities, but
with semantics that may vary across implementations and user practices. 

\subsubsection*{Decentralized Identifiers (DIDs)}
Decentralized Identifiers (DIDs) provide globally unique identifiers whose
associated metadata is obtained through a resolution process. Resolution
produces a DID Document specifying public keys, metadata, service endpoints,
and key rotation procedures. DID Documents are retrieved via method-specific
resolvers that may depend on blockchains or other registries. DIDs illustrate an
approach in which identifiers are decoupled from cryptographic keys and require
external infrastructure for interpretation. \cite{W3C2022DIDs}

\subsubsection*{Ledger-Based Identity Models}
Blockchain and distributed-ledger systems provide tamper-evident publication
and global ordering of identity updates or credential assertions by maintaining
an append-only, consensus-maintained log governed by network-wide consensus
mechanisms. \cite{EZAWA2023100126} These systems depend on consensus
infrastructure to establish shared state that multiple parties can reference.
They show how trust anchoring and credential visibility can be decentralized
via coordination across a global network.

\subsection{Delegation and Authorization Frameworks} 

\subsubsection*{SPKI/SDSI}
SPKI and SDSI introduced key-centered naming, explicit authorization
certificates, and locally meaningful identity scopes. SDSI emphasized local
namespace semantics and linked local names, while SPKI defined certificate
formats and operational mechanisms for expressing authorization and delegation
directly in terms of keys. \cite{Rivest2001SPKI} These systems explored
certificate-based trust graphs, non-hierarchical naming, and the challenges of
revocation and freshness in decentralized contexts.

\subsubsection*{Macaroons}
Macaroons represent authorization through chained caveats that progressively
narrow permitted actions as a token is attenuated. They use nested HMAC-based
constructions and are designed for efficient deployment in web and cloud
services. \cite{Birgisson2014Macaroons} Caveat verification may rely on shared
secrets or online checks, and macaroons show how capabilities can be refined
through incremental constraint embedding within bearer tokens.

\subsubsection*{Biscuit Authorization Tokens}
Biscuit tokens embed a sequence of signed blocks, enabling expressive delegated
authorization with offline verification. Biscuit combines public-key signatures
with a small logic language to represent facts, rules, and checks carried
directly within a token. \cite{BiscuitSecDocs} This design demonstrates how
token-carried logic and chained signatures can encode authorization properties
in a form that can be evaluated locally by verifiers.

\subsubsection*{Capability-Oriented Composition}
Miller's \emph{Robust Composition} describes a capability-based object model in
which authority is conveyed through possession of references, and system
interaction is constrained by controlling reference propagation and use.
The work treats access control and concurrency control as related problems of
managing inter-object causality in distributed systems. \cite{Miller2006}

The model is developed using asynchronous message passing, promise references,
and explicit handling of partial failure, with examples drawn from the E
programming language and distributed object systems.

\subsection{Summary of Conceptual Positioning} 
Across these systems, several recurring ideas appear: identity may be
institutionally assigned (PKI), socially constructed (PGP), resolver-mediated
(DIDs), or ledger-anchored; authorization may be represented through
certificates (SPKI/SDSI), chained attenuation (Macaroons), logic-bearing tokens
(Biscuit), or capability possession (capability systems). The ZI-CG model
adopts assumptions that differ from the above and is discussed formally
later in the paper. The purpose of this section is to establish the conceptual
landscape and terminology upon which ZI-CG builds.

\section{Zero-Infrastructure Capability Graphs (General Model)} 

\subsection{Model Overview} 
The Zero-Infrastructure Capability Graph (ZI-CG) model provides a framework for
identity and trust that remains correct and verifiable without any reliance on
online infrastructure. Traditional systems depend on resolvers, directories,
revocation services, or trusted authorities. In contrast, a ZI-CG system
represents all trust-relevant information as a finite set of self-contained,
cryptographically signed statements. A verifier resolves identity, delegation,
and revocation using only this local token set, with no external lookup or
shared global state.

The essential insight behind ZI-CG is conceptual separation: trust evaluation
must not depend on service or network availability. Many traditional systems
combine these concerns by requiring that trust evaluators retrieve information
from online services. A ZI-CG system avoids this entanglement by ensuring that
identity binding, capability propagation, and state changes such as revocation
or burn are fully encoded within the signed statements themselves. Trust
becomes a matter of computation rather than communication.

A ZI-CG system therefore has two main components:

\begin{enumerate}
  \item A set of immutable, self-verifying tokens that encode identity and trust
relationships using signed and content-addressed data.

  \item A deterministic evaluation procedure that interprets these tokens to
determine identity authenticity, delegated capability, and revocation status
entirely offline.

\end{enumerate}

The model does not prescribe a specific token format, identity string, or
message encoding. Instead, it defines the structural constraints that any
zero-infrastructure trust system must satisfy in order to ensure that verifiers
operating in disconnected or adversarial environments will always reach
consistent trust judgments.

Together, these elements describe the conceptual structure of a ZI-CG system.
The next sections formalize the requirements that follow from this structure
and define the token semantics and evaluation rules that allow the model to
operate deterministically in offline environments.

\subsection{Fundamental Goals} 
The ZI-CG model is guided by two foundational goals that shape its structure
and constraints.

\subsubsection*{Goal 1: A secure, cryptographic substrate for identity and trust that works entirely offline}
A ZI-CG system must enable verifiers to authenticate identities, validate
statements, determine delegated authority, and detect revocation using only a
locally provided set of signed tokens. No network reachability or supporting
infrastructure can be assumed. Correctness must hold across network partitions,
constrained environments, and adversarial conditions.

\subsubsection*{Goal 2: Clean separation of trust evaluation from data availability}
Existing identity systems often bind trust decisions to data retrieval. For
example, resolving an identity, checking a key, fetching a revocation status,
or consulting a registry all require active infrastructure. ZI-CG separates
these concerns. All information needed for trust evaluation is embedded in the
tokens themselves, allowing data distribution to be handled independently by
any suitable mechanism. This could be synchronous, asynchronous, centralized,
or peer to peer. None of these choices affect the correctness of trust
computation when the relevant tokens are available to the verifier.

These goals motivate the model's design constraints. The following sections
formalize the requirements a system must satisfy in order to qualify as a ZI-CG
system and describe the token structures and evaluation semantics that
follow from those constraints.

\subsection{Design Constraints and Requirements} 
\label{sec:requirements}

\subsubsection*{Requirement 1: Self-Resolving Identity}
A ZI-CG system requires an identity format that can be validated using only the
identifier itself and the public key it refers to. No lookup, resolver,
metadata document, or external authority may be required to determine whether
an identity correctly binds to a key. Identity verification must remain
possible for any verifier operating offline or in an adversarial environment.
This requires the identifier to encode enough structure for the verifier to
confirm its binding to a key through computation alone. Without self-resolving
identity, identity validation would require online lookup or external
infrastructure, which violates the zero-infrastructure constraint.

\subsubsection*{Requirement 2: Local State Resolution by Proof-of-Omission}
All state changes relevant to trust evaluation, such as revocation or identity
termination, must be represented as signed tokens within the local token set. A
verifier determines effective state by omitting tokens that have been
invalidated, not by consulting external lists or status services. Since a ZI-CG
verifier cannot rely on network connectivity or shared mutable state, all
revocation information must be expressible and verifiable using only signed data
available locally. Without local, data-driven state resolution, revocation
would require retrieval from an external authority, which violates the
zero-infrastructure constraint.

\subsubsection*{Requirement 3: Attenuated Delegation}
Delegation of authority must always reduce or preserve capability, never expand
it. Each delegation token must specify the scope it endorses, and the effective
capability of any delegation chain must be the intersection of the scopes along
that chain. In a ZI-CG system, delegation cannot rely on policy engines,
external authorization services, or online context. Delegated authority must be
encoded and bounded directly in the signed statements themselves to guarantee
predictable and safe propagation. Without attenuated delegation, capability
could expand in ways that require external policy or infrastructure to
constrain it, which violates the zero-infrastructure constraint.

\subsubsection*{Requirement 4: Structural Determinism and Offline Evaluation}
Trust evaluation must be a deterministic function of the provided token set and
the verifier's chosen trust roots. Two verifiers given the same inputs must
produce the same trust result, regardless of environment or timing. Determinism
ensures that evaluation does not depend on network state, shared global state,
or other externally mediated factors. This property allows disconnected
verifiers to reach consistent conclusions based solely on the signed statements
they possess. Without structural determinism, trust evaluation would require
external coordination to resolve inconsistencies, which violates the
zero-infrastructure constraint.

These requirements represent the minimum structural conditions for any ZI-CG
trust system. Specific systems may impose additional constraints, but the model
requires at least these four.

\subsection{Self-Verifying Trust Statements} 
All trust assertions in a ZI-CG system are represented as self-verifying
statements. Each statement is encoded as a signed token that contains all
information needed for a verifier to confirm its origin and integrity. The
token includes the issuer's public key, and the issuer's identifier is
cryptographically bound to that key by a deterministic, collision-resistant
function. The token is signed using the corresponding private key. A verifier
can therefore determine, without any external lookup, that the statement could
only have been produced by the holder of that private key and that its contents
have not been altered.

\subsection{Informal Model Description} 
Given a set of self-verifying trust statements, a ZI-CG verifier evaluates
authorization by reasoning solely over the signed tokens available to it. No
external services, authoritative registries, or online resolution mechanisms
are consulted. The token set presented to the verifier therefore defines the
entire universe of trust-relevant information for that evaluation. Any signed
statement that could influence the trust decision must therefore appear within
this token set.

The model defines four categories of tokens.  An \textbf{attestation} records a
statement of fact asserted by its issuer. A \textbf{vouch} expresses delegated
trust by endorsing another token and optionally restricting the scope of that
endorsement. A \textbf{revocation} retracts a specific earlier statement by
invalidating the referenced token. A \textbf{burn} token signals that the
issuer's identity is no longer valid and that none of its statements should be
accepted from that point forward. These four token types represent the minimal
categories required for identity, delegation, and revocation in the basic
model, but the framework does not restrict systems from defining additional
token types for other purposes. The essential requirement is that any statement
that matters for trust evaluation must be expressible as a self-verifying token
available to the verifier.

The ZI-CG model does not assign trust directly to identities. Instead, trust is
attached to specific signed statements an issuer has made. A vouch token
therefore does not mean that Alice is trusted in general. It means that Alice
has issued a concrete, verifiable statement asserting trust in another
statement for a particular purpose. This shift from identity-centric to
statement-centric trust is central to how capability propagation and revocation
behave in ZI-CG systems.

A verifier evaluates a token set by constructing a graph that reflects these
relationships. Each token set is first validated and examined to collect any
revocation or burn statements. For each token, the verifier then applies a
proof-of-omission check. A token is added as a node in the graph if and only if the
revocation set does not contain a revocation that targets it and the burn set
does not contain a burn that targets its issuer. This graph structure arises
naturally from the content-addressed references within the tokens, since these
references define how statements depend on and endorse one another. This
procedure ensures that the graph includes only those statements whose validity
is not contradicted by the information present in the token set. The directed
graph that results represents the verifier's unified view of trust, since
membership in the graph confirms that no revocation or burn relevant to that
node is present. With the valid token graph established, the verifier next
determines which capabilities are supported by the endorsed relationships it
encodes.

Once this graph has been constructed, capabilities propagate along chains of
vouch tokens. The effective capability associated with any chain is determined
by intersecting the scopes that appear along the path. This propagation occurs
entirely within the graph and relies only on the signed tokens that remain
after state resolution. The result is a deterministic evaluation of trust that
depends only on the content of the token set. Identical token sets always yield
identical trust outcomes, regardless of where or when the evaluation occurs.

A ZI-CG system is naturally actor-centric. In most deployments the party
seeking to exercise a capability presents the signed tokens that justify it. A
verifier may also maintain its own locally obtained tokens, such as revocations
or burns that reflect organizational policy or known compromise. When
evaluating a request, the verifier considers the combined set of tokens
supplied by the actor and those it holds locally. The evaluation process treats
this combined set exactly the same as any other token set, and its correctness
depends only on the signed statements available to the verifier, not on how
those statements were obtained.

The next section provides the formal semantics that give precise meaning to the
token structures and evaluation rules outlined above.

\section{Formal Model} 
\label{sec:formal-model}

\subsection{Identities} 
\label{sec:identities}
An identity in a ZI-CG system is a string that can be validated using only the
identity itself and the public key associated with it. Let (I) denote the set
of identity strings and (PK) denote the set of public keys. Let ($\mathcal{M}$)
be the space of optional descriptive or human-readable components, and let
($\mathcal{H}$) be the space of collision-resistant digests.

A ZI-CG system defines two deterministic functions

\begin{equation}
\mathsf{parse} : I \rightarrow \mathcal{M} \times \mathcal{H}
\end{equation}
\begin{equation}
\mathsf{id} : PK \times \mathcal{M} \rightarrow I
\end{equation}

and a collision-resistant function

\begin{equation}
h : PK \rightarrow \mathcal{H}.
\end{equation}

Intuitively, ($\mathsf{id}$) constructs an identity string from a public key and
optional metadata, while ($\mathsf{parse}$) recovers the metadata and a stored
digest from that string.

The system must satisfy the following consistency property. For all $(pk \in PK) and (m \in \mathcal{M})$, let

\begin{equation}
i = \mathsf{id}(pk, m) \qquad \mathsf{parse}(i) = (m', d).
\end{equation}

Then it must hold that $m' = m$ and

\begin{equation}
d = h(pk).
\end{equation}

Verification of the binding between an identity $i$ and a public key $pk$
proceeds by computing $\mathsf{parse}(i) = (m, d)$, recomputing $d' = h(pk)$,
and accepting the binding if and only if $d' = d$. 

Identities may include arbitrary descriptive material in $m$, but all such material
must be incorporated into the identity string in a way that preserves this
self-verifying property. Within the model, principals are identified by their
full identity strings, which include both any descriptive metadata and the
key-derived digest; two principals are considered the same if and only if their
identity strings are identical. \textbf{Consequently, the same public key combined
with different descriptive material yields distinct identity strings and
therefore distinct principals within the model.}

Because $h$ is collision-resistant and $\mathsf{id}$ is deterministic, it is
computationally infeasible for two distinct public keys to produce the same
digest component in an identity, and therefore infeasible to construct two
distinct principals that share the same identity string without breaking the
underlying hash function or encoding.

This self-resolving construction ensures that identity validation remains
possible without external lookups, and that any reference to an identity in a
token can be interpreted correctly using only information available to the
verifier in the token itself.

\subsection{Tokens} 
\label{sec:tokens}

A ZI-CG system encodes all trust-relevant information as
\textit{self-verifying, content-addressed tokens that are independently
authenticatable}.  

A key property of this construction is that each token constitutes a
\textit{self-authenticating statement}. Unlike conventional signed objects,
which rely on external mechanisms to establish the validity of the signing key,
a ZI-CG token carries all information required to verify both the authenticity
of the statement and the cryptographic identity of its issuer. The issuer
identifier, the verification key material, and the signature are co-resident
within the token body, allowing a verifier to authenticate the statement using
only the token itself.

This design deliberately collapses the distinction between identity verification
and statement verification. In typical systems, verifying a signature establishes
only that some key signed the data, while determining whether that key represents
a particular identity requires external resolution through certificates,
directories, or resolvers. In contrast, a ZI-CG token is structured so that
successful signature verification simultaneously establishes the binding between
the issuer identifier and the signing key. As a result, token authenticity becomes
a property of the data alone.

Tokens are defined in three layers: the signature layer, the universal body
layer, and the payload layer.

\subsubsection{Signature Layer}

Let $T$ denote the set of all tokens. Each token $t \in T$ has the form

\begin{equation}
t = \bigl(\mathsf{body}(t), \mathsf{sig}(t)\bigr)
\end{equation}

where $\mathsf{body}(t)$ is a structured object representing the semantic
content of the token, and $\mathsf{sig}(t)$ is a digital signature.

Let

\begin{equation}
\mathsf{serialize}(\mathsf{body}(t)) \in \{0,1\}^\ast
\end{equation}

be a canonical encoding of the token body. The signature is computed over this
encoding using the issuer's private key.

To uniquely bind both the token's content and its issuer, the
content-addressed identifier of the token is defined as

\begin{equation}
\mathsf{tid}(t) =
H\bigl(\mathsf{serialize}(\mathsf{body}(t)) \,\|\, \mathsf{sig}(t)\bigr)
\end{equation}

where $H$ is a collision-resistant hash function and $\|$ denotes byte
concatenation. The value $\mathsf{tid}(t)$ serves as the canonical identifier
of token $t$ within the model.

This construction ensures that any change to the token's body or signature
produces a distinct identifier. As a result, a token's meaning and authorship
are cryptographically bound into a single immutable artifact: the token cannot
be altered, re-signed by a different identity, or reinterpreted without
changing its identifier and invalidating all references to it.

\subsubsection{Universal Body Structure} 
The token body contains the issuer's identity string, the issuer's public key,
and a structured payload:

\begin{equation}
\mathsf{body}(t) = \bigl(\mathsf{iss}(t), \mathsf{pk}(t), \mathsf{payload}(t)\bigr)
\end{equation}

\begin{itemize}
  \item $\mathsf{iss}(t) \in I$ is the issuer identity string.
  \item $\mathsf{pk}(t) \in PK$ is the issuer's public key.
  \item $\mathsf{payload}(t)$ carries the semantic content of the token, defined below.
\end{itemize}

Both $\mathsf{iss}(t)$ and $\mathsf{pk}(t)$ are included inside the signed
body. This prevents attackers from substituting a different identity or public
key without changing the token identifier and breaking the signature.

Token validity has two components:

\textbf{Signature validity}

\begin{equation}
\mathsf{sigvalid}(t) \iff
\mathsf{verify\_sig}\bigl(\mathsf{pk}(t), \mathsf{serialize}(\mathsf{body}(t)), \mathsf{sig}(t)\bigr)
\end{equation}

\textbf{Identity binding validity}

Using the identity mechanism from Section \ref{sec:identities}:

\begin{equation}
\mathsf{idvalid}(t) \iff
\bigl(\mathsf{parse}(\mathsf{iss}(t)) = (m, d)\bigr) \ \land\ \bigl(d = h(\mathsf{pk}(t))\bigr)
\end{equation}

\pagebreak[1]
\textbf{Overall token validity}

\begin{equation}
\mathsf{valid}(t) \iff \mathsf{sigvalid}(t) \land \mathsf{idvalid}(t)
\end{equation}

A verifier must reject any token that fails either condition. These predicates
ensure that every accepted token is both correctly signed and correctly bound
to its stated issuer identity.

\subsubsection{Payload Structure} 
The payload describes what the issuer is asserting. It has a common shape with
three components:

\begin{equation}
\mathsf{payload}(t) = \bigl(\mathsf{kind}(t), \mathsf{typefields}(t), \mathsf{extra}(t)\bigr).
\end{equation}

Every token has a \textbf{kind} that determines how the token participates in the capability graph and which additional fields are required:

\begin{equation}
\mathsf{kind}(t) \in {\text{attest}, \text{vouch}, \text{revoke}, \text{burn}}.
\end{equation}

The $\mathsf{typefields}(t)$ component contains \textbf{type-specific fields}
that depend on the kind of the token. At the model level, ZI-CG requires the
following minimum structure:

\begin{itemize}
    \item If $\mathsf{kind}(t) = \text{attest}$, then
    \begin{equation}
      \mathsf{typefields}(t) = \emptyset.
    \end{equation}

The token carries only assertions in $\mathsf{extra}(t)$.

  \item If $\mathsf{kind}(t) = \text{vouch}$, then

\begin{equation}
  \mathsf{typefields}(t) = \bigl(\mathsf{ref}(t), \mathsf{scope}(t)\bigr)
\end{equation}

where $\mathsf{ref}(t)$ is the content-addressed identifier $\mathsf{tid}(t')$ of
the referenced token $t'$, and $\mathsf{scope}(t)$ encodes the attenuation applied to that
endorsement.

  \item If $\mathsf{kind}(t) = \text{revoke}$, then
\begin{equation}
\mathsf{typefields}(t) =
\bigl(\mathsf{revoke\_target}(t), \mathsf{ref\_subj}(t)\bigr)
\end{equation}

where:
\begin{itemize}
  \item $\mathsf{revoke\_target}(t)$ is a deterministic selector that uniquely
  identifies a prior statement issued by the same issuer,
  \item $\mathsf{ref\_subj}(t)$ is the subject referenced by that prior statement.
\end{itemize}

  \item If $\mathsf{kind}(t) = \text{burn}$, then

\begin{equation}
\mathsf{typefields}(t) = \bigl(\mathsf{ref\_id}(t)\bigr)
\end{equation}

where $\mathsf{ref\_id}(t) \in I$ MUST equal $\mathsf{iss}(t)$.

\end{itemize}

Subsequent sections formalize how $\mathsf{ref}(t)$, $\mathsf{scope}(t)$, and
the kind interact to produce the capability graph and state transitions.

The $\mathsf{extra}(t)$ component represents arbitrary application-specific \textbf{additional fields} or metadata. These fields are:

\begin{itemize}
  \item fully covered by the signature,
  \item included in the content hash $\mathsf{tid}(t)$, and
  \item available to applications for interpretation,
\end{itemize}

but they are \textit{ignored} by the core ZI-CG evaluation semantics. The
evaluator's behavior is defined entirely in terms of

\begin{equation}
\mathsf{iss}(t),\ \mathsf{pk}(t),\ \mathsf{kind}(t),\ \mathsf{ref}(t),\ \mathsf{scope}(t),\ \mathsf{tid}(t)
\end{equation}

and the token's validity predicate. This allows systems to extend token contents without changing the underlying trust model.

This definition of tokens provides a universal, self-verifying substrate on
which the remainder of the ZI-CG model is built. The next subsections use these
structures to define token sets, state resolution by proof of omission, and the
construction of the capability graph.

\subsection{Token type semantics} 
\label{sec:token-type-semantics}

Given the general token structure from Section \ref{sec:tokens}, this section specifies the
semantics of the four token kinds that define the ZI-CG model: attestation,
vouch, revocation, and burn. These semantics describe how each token type
contributes to state, delegation, and eventual capability evaluation. 

Let $T_{\text{valid}}$ denote the set of tokens that satisfy
$\mathsf{valid}(t)$ as defined in Section \ref{sec:tokens}. Several descriptions below refer
informally to a cleaned token set $T_{\text{clean}}$, which consists of the
valid tokens that remain after revocation and burn semantics are applied. 

Section \ref{sec:state-resolution} provides the formal definition of
$T_{\text{clean}}$ and the proof-of-omission rules that construct it.

\subsubsection{Attestation tokens} 
An attestation token has

\begin{equation}
\mathsf{kind}(t) = \text{attest} \qquad \mathsf{typefields}(t) = \emptyset
\end{equation}

Its payload consists only of $\mathsf{extra}(t)$, which encodes statements of
fact asserted by the issuer. The ZI-CG model does not impose intrinsic
structure or truth conditions on these statements. Instead, it treats them as
application-level claims that may be accepted or rejected depending on whether
the token itself is trusted by the evaluator.

Formally, attestation tokens

\begin{itemize}
  \item do not introduce edges in the capability graph,
  \item do not change revocation or burn state, and
  \item may be the target of vouch or revocation tokens.
\end{itemize}

Although attestation tokens do not by themselves induce delegation edges or
capability propagation, their self-authenticating structure makes them useful
even in isolation. An attestation token constitutes a verifiable statement of
fact whose origin can be established directly from the token contents, without
reliance on directories, certificates, or external resolution. In applications
where the issuer is already known or accepted by policy, such tokens can be
consumed directly as authenticated assertions, independent of any additional
vouching or graph-based trust composition.

The ZI-CG model treats attestation as a minimal primitive for authenticated
claims, upon which richer trust semantics such as delegation, attenuation, and
revocation may be layered. However, the correctness and verifiability of an
attestation token does not depend on the presence of those mechanisms, only on
the cryptographic self-authentication of the statement itself.

An attestation token $t$ contributes its asserted facts only if it is present
in the cleaned token set $T_{\text{clean}}$ and is reachable from a trust
root under the delegation semantics defined later. The core ZI-CG model is
agnostic to the internal structure of $\mathsf{extra}(t)$; it is sufficient
that the application can interpret these claims once the token has been deemed
trusted.

\subsubsection{Vouch tokens}
\label{sec:vouch-tokens}

A vouch token represents a delegated endorsement of another token. It has

\begin{equation}
\mathsf{kind}(t) = \text{vouch} \qquad
\mathsf{typefields}(t) = \bigl(\mathsf{ref}(t), \mathsf{scope}(t)\bigr)
\end{equation}

Here:

\begin{itemize}
  \item $\mathsf{ref}(t)$ carries the token identifier of the endorsed token:
\begin{equation}
  \mathsf{ref}(t) = \mathsf{tid}(u)
\end{equation}
for some token $u \in T_{\text{valid}}$ with
$\mathsf{kind}(u) \in \{\text{attest}, \text{vouch}\}$.

  \item $\mathsf{scope}(t)$ is a \textbf{set of capabilities},
  $\mathsf{scope}(t) \subseteq S$, which constrains the delegation.
  A vouch may therefore authorize multiple capabilities at once, but all of
  them are subject to attenuation when propagated along a path.
\end{itemize}

Informally, a vouch token $t$ states that the issuer $\mathsf{iss}(t)$ accepts
the claims or authority expressed by the referenced token $u$ within the limits
described by $\mathsf{scope}(t)$. Vouch tokens that remain after revocation and
burn semantics are applied induce directed edges in the capability graph. These
edges represent the propagation of attenuated authority from the issuer of the
vouch toward the subject of the referenced token.

Vouch tokens may reference only attestation or vouch tokens. References to
tokens of kind $\text{revoke}$ or $\text{burn}$ are ignored by the ZI-CG
evaluation semantics, as such tokens do not represent delegable authority.

If $\mathsf{ref}(t)$ does not match the identifier of any token that survives
state resolution, the vouch has no effect on the graph. Such vouch tokens are
syntactically valid but semantically inert with respect to capability
propagation.

\subsubsection{Revocation tokens} 
\label{sec:revocation-tokens}
A revocation token retracts a specific earlier statement or endorsement made by
its issuer. It has

\begin{equation}
\mathsf{kind}(t) = \text{revoke}
\end{equation}

and its type-specific fields identify both the prior statement being withdrawn
and the subject of that prior statement. At the model level,

\begin{equation}
\mathsf{typefields}(t) =
\bigl(\mathsf{revoke\_target}(t), \mathsf{ref\_subject}(t)\bigr).
\end{equation}

Informally, a revocation token states:

"I, $\mathsf{iss}(t)$, no longer stand behind the earlier statement identified by
$\mathsf{revoke\_target}(t)$ concerning subject $\mathsf{ref\_subject}(t)$."

The two fields are required together so that the revocation unambiguously
identifies a single prior act by the issuer and cannot be redirected to a
different subject by substituting or modifying the targeted statement.

A revocation token $r$ targets a token $u$ if and only if the following
conditions hold:

\begin{enumerate}
  \item $u$ is an attestation or vouch:
  \begin{equation}
    \mathsf{kind}(u) \in \{\text{attest}, \text{vouch}\}
  \end{equation}

  \item the issuer matches:
  \begin{equation}
    \mathsf{iss}(r) = \mathsf{iss}(u)
  \end{equation}

  \item the revocation target matches $u$:
  \begin{equation}
    \mathsf{matches\_target}(r, u)
  \end{equation}

  \item the subject matches:
  \begin{equation}
    \mathsf{ref\_subject}(r) = \mathsf{ref\_subject}(u)
  \end{equation}
\end{enumerate}

The predicate $\mathsf{matches\_target}(r, u)$ holds if and only if the value
$\mathsf{revoke\_target}(r)$ deterministically identifies token $u$ among all
tokens issued by $\mathsf{iss}(r)$ in the supplied token set.

The model requires that at most one token $u$ satisfies
$\mathsf{matches\_target}(r, u)$ for any given revocation token $r$.

For attestation tokens, the statement itself is treated as its own subject.

The effect of revocation is to remove the targeted statement from the effective
state. When the cleaned token set $T_{\text{clean}}$ is constructed, any
statement token $u$ for which there exists a revocation token $r$ satisfying the
predicate above is excluded from $T_{\text{clean}}$. Subject tokens referenced
by those statements are not removed by this mechanism; the revocation
eliminates only the issuer's prior attestation or endorsement. Other parties may
continue to issue their own statements about the same subject.

The revocation tokens themselves remain part of $T_{\text{clean}}$ (subject to
their own validity and any applicable burn semantics), ensuring their effects
are visible to any evaluator that receives the same token set.

To maintain simplicity and ensure the capability graph remains acyclic,
revocation tokens do not target other revocation tokens or burn tokens.
Revocation is monotonic: adding a revocation token can only remove attestation
and vouch tokens issued by the revoker, thereby reducing the available
delegation paths or accepted statements; it never reinstates authority that has
already been removed.

\subsubsection{Burn tokens} 
\label{sec:burn-tokens}
A burn token terminates an identity within the model. It has

\begin{equation}
\mathsf{kind}(t) = \text{burn} \qquad
\mathsf{typefields}(t) = \bigl(\mathsf{burn}(t)\bigr),
\end{equation}

where $\mathsf{burn}(t)$ denotes the identity that is to be treated as no longer valid.

To ensure a burn token only terminates its own identity, we require that the
identity referenced by the burn token must be the identity of the burn token
itself:

\begin{equation}
\mathsf{burn}(t) = \mathsf{iss}(t).
\end{equation}

Given $T_{\text{valid}}$, a burn token $b$ targets all valid tokens whose issuer matches the burned identity. For any token $u \in T_{\text{valid}}$,

\begin{equation}
\mathsf{burned\_by}(u, b) \iff \mathsf{burn}(b) = \mathsf{iss}(u).
\end{equation}

When the cleaned token set $T_{\text{clean}}$ is constructed, any token $u$
that is burned by at least one burn token is excluded. The burn tokens
themselves remain in $T_{\text{clean}}$ (unless invalid or themselves removed
by a more restrictive policy) so that their effect persists wherever they are
present.

Burn semantics are stronger than revocation. A revocation applies only to a
single referenced statement, while a burn applies to all statements ever issued
by the burned identity, including attestations, vouches, and revocations. As
with revocation, the effect of a burn is monotonic. Adding a burn token can
only remove authority from the effective state and cannot be undone by any
other token within the model.

For simplicity of reasoning and to preserve the intended meaning of burn as
identity termination, the model assumes that burn tokens do not target other
burn tokens. Multiple burn tokens for the same identity are allowed but
redundant; they do not alter the resulting state beyond the effect of a single
burn.

These token-type semantics describe how individual tokens influence the
eventual effective state of the system. The next section formalizes token sets
and the proof-of-omission procedure that combines revocation and burn semantics
to construct the cleaned token set $T_{\text{clean}}$, which serves as the
basis for capability graph construction and delegation evaluation.

\subsection{State resolution and construction of $T_{\text{clean}}$} 
\label{sec:state-resolution}
State resolution is the process by which a verifier removes tokens whose
effects have been negated by revocation or identity termination. The result is
the cleaned token set $T_{\text{clean}}$, which is the only token set used
for capability graph construction and evaluation. All state resolution is
purely local and depends only on the tokens supplied to the evaluator.

\subsubsection{Input token set and application-level filtering} 
Let $T$ be the set of tokens supplied to the evaluator.

The ZI-CG model intentionally does not prescribe how applications obtain $T$ or
which tokens they choose to include. Applications may construct $T$ from one or
more actors or data sources and may apply any filtering policy they require.
Typical examples include:

\begin{itemize}
  \item temporal validity windows,
  \item contextual relevance,
  \item policy-driven exclusion,
  \item size or scope limits.
\end{itemize}

These filtering decisions occur entirely \textbf{before} the ZI-CG evaluator is
invoked. The model assumes that all tokens present in $T$ are intended to
participate in trust computation, and that tokens not appearing in $T$ are
irrelevant to the evaluation.

Applications that possess a reliable local notion of time may optionally
interpret token expiration fields as part of their filtering logic. Expiration
does not alter ZI-CG semantics; it simply determines whether a token is
included in $T$. When used, expiration functions as a natural attenuation
mechanism: delegations and attestations automatically phase out as their
issuer-defined validity windows close. In practical deployments, this can
substantially reduce the need for explicit revocation, since stale
authorizations are excluded from $T$ prior to evaluation. Importantly,
disregarding expiration cannot expand effective authority within the model,
while honoring it may remove tokens that would otherwise have contributed to
the trust graph. Thus applications may employ time-based filtering to
strengthen operational security without introducing any dependency into the
ZI-CG core model.

\subsubsection{Valid token set} 
The valid token set $T_{\text{valid}}$ consists of all tokens in $T$ that satisfy:

\begin{itemize}
  \item structural validity,
  \item correct signature verification using the embedded public key,
  \item correct identity binding between issuer identifier and public key.
\end{itemize}

Formally,

\begin{equation}
T_{\text{valid}} = \{t \in T \mid \mathsf{valid}(t) \}
\end{equation}

\subsubsection{Burn predicate}
\label{sec:burn-predicate}

A burn token indicates that all statements issued by a particular identity must
be disregarded.

For any token $u \in T_{\text{valid}}$, define:

\begin{equation}
\mathsf{burned}(u) \iff
\mathsf{kind}(u) \neq \text{burn} \ \land\ 
\exists b \in T_{\mathrm{valid}} \text{ such that }
\mathsf{kind}(b) = \text{burn} \land \mathsf{burn}(b) = \mathsf{iss}(u).
\end{equation}

A token $u$ is burned if there exists a valid burn token whose referenced
identity matches the issuer of $u$. Burn semantics affect all non-burn token
kinds, including attestations, vouches, and revocations. Burn tokens themselves
are not removed by this predicate, so that their effect remains visible wherever
they are present.

\subsubsection{Revocation predicate} 
A revocation token retracts a specific attestation or vouch previously issued by the same identity.

Using the matching rule defined in Section \ref{sec:revocation-tokens}, define:

\begin{equation}
\mathsf{revoked}(u) \iff
\exists r \in T_{\text{valid}} \text{ such that } \mathsf{revokes}(r, u)
\end{equation}

where $\mathsf{revokes}(r, u)$ holds if and only if the targeting conditions
defined in Section \ref{sec:revocation-tokens} are satisfied.

Revocation does not apply to burn tokens or other control tokens, and it does
not directly remove subject tokens issued by other identities. It removes only
the specific statements that the revoking issuer previously made.

\subsubsection{Cleaned token set} 
The cleaned token set is defined by removing all burned and revoked tokens from $T_{\text{valid}}$:

\begin{equation}
T_{\text{clean}} =
\{u \in T_{\text{valid} }
\mid
\neg \mathsf{burned}(u)
\land
\neg \mathsf{revoked}(u)
\}
\end{equation}

In other words:

\begin{itemize}
  \item tokens issued by burned identities are discarded,
  \item attestations and vouches explicitly revoked by their issuer are discarded,
  \item all other valid tokens remain.
\end{itemize}

Burn and revocation semantics are monotonic. Adding such tokens to $T$ can only remove additional tokens from $T_{\text{clean}}$; it can never restore a token that has already been removed. The resulting set $T_{\text{clean}}$ represents the verifier's complete and self-contained view of the current state of all trust-relevant statements within the provided token set. It forms the basis for capability graph construction and delegation semantics in the next section.

\subsection{Capability graph construction} 
\label{sec:capability-graph-construction}
Given the cleaned token set $T_{\text{clean}}$, the ZI-CG model represents delegation structure as a directed graph whose nodes are delegable statements and whose edges are induced by surviving vouch tokens. This graph provides the structural basis for capability evaluation in the next section. Graph construction is fully deterministic and depends only on $T_{\text{clean}}$.

\subsubsection{Nodes} 
The node set of the capability graph consists of all delegable statements that
remain after state resolution. Formally,

\begin{equation}
V = \{ u \in T_{\text{clean}} \mid \mathsf{kind}(u) \in \{\text{attest}, \text{vouch}\} \}.
\end{equation}

Each node $u \in V$ represents a trust statement that may participate in
delegation and capability propagation. Control tokens, including revocation and
burn tokens, influence the construction of $T_{\text{clean}}$ but do not appear
as nodes in the capability graph.

\subsubsection{Edges from vouch tokens} 
Directed edges arise only from vouch tokens in $T_{\text{clean}}$. Intuitively, a vouch token $v$ induces an edge representing delegation from its issuer toward the referenced statement, provided that token also survives state resolution.

Formally, the edge set $E \subseteq V \times V$ is defined as

\begin{equation}
E = \{ (v, u) \mid
v \in V,
\ \mathsf{kind}(v) = \text{vouch},
\ \mathsf{ref}(v) = \mathsf{tid}(u),
\ u \in V
\}
\end{equation}

If a vouch token $v$ refers to a token that does not appear in
$T_{\text{clean}}$, no edge is created. In that case $v$ remains as an
isolated node. This reflects the fact that an endorsement of a statement that
has been removed or never provided cannot influence delegation.

\subsubsection{Scope labels on edges} 
Each vouch token carries a scope value that constrains the authority it
conveys. Let $S$ denote the scope space and let $\mathsf{scope}(v) \in S$ be
the scope associated with a vouch token $v$.

For every edge $(v, u) \in E$ that arises from a vouch token $v$, the edge is annotated with the scope of the vouch:

\begin{equation}
\mathsf{scope}(v, u) = \mathsf{scope}(v).
\end{equation}

These scope labels are not interpreted during graph construction. They are used
in the evaluation phase to compute the effective capability associated with a
path by intersecting the scopes of the edges along that path. Graph
construction therefore records the structural and scoping information required
for later capability evaluation, but does not perform any policy reasoning
itself.

\subsubsection{Resulting capability graph} 
The capability graph associated with a cleaned token set $T_{\text{clean}}$ is the pair

\begin{equation}
G(T_{\text{clean}}) = (V, E),
\end{equation}

where $V$ and $E$ are defined as above, and each edge $(v, u) \in E$ carries a scope label $\mathsf{scope}(v, u)$.

This construction has several important properties.

\begin{itemize}
  \item \textbf{Locality.} Graph structure is determined entirely by $T_{\text{clean}}$.
No external state, resolver, or registry is consulted.

  \item \textbf{Determinism.} For any fixed $T_{\text{clean}}$, the graph
$G(T_{\text{clean}})$ is uniquely determined. Two evaluators given the same
cleaned token set will construct identical graphs.

  \item \textbf{Monotonicity under removal.} If tokens are removed from
$T_{\text{clean}}$, the resulting graph is obtained by removing the
corresponding nodes and any incident edges. No new edges are introduced by
removing tokens.

  \item \textbf{Acyclic structure.} Because vouch tokens reference other tokens by
their content-addressed identifiers, a token may only reference statements that
existed prior to its creation. This induces a strict partial order over tokens
and guarantees that the resulting capability graph is acyclic. Revocation and
burn semantics only remove nodes or edges and therefore preserve acyclicity.
The evaluation procedure in the next section can therefore treat delegation as
propagation along finite directed paths without requiring external cycle
detection mechanisms.

\end{itemize}

The capability graph $G(T_{\text{clean}})$ captures all potential delegation
and endorsement structure that survives state resolution. The next section
defines how this graph, together with a verifier's chosen trust roots, is used
to compute effective capabilities for particular tokens and scopes.

\subsection{Capability evaluation} 
\label{sec:capability-evaluation}

Capability evaluation determines whether a specific token $t$ is authorized for
a requested scope of authority, given a cleaned token set $T_{\text{clean}}$
and a verifier's set of trusted principals and their associated scopes.
Evaluation is a local computation that begins at the token under evaluation and
searches backward through delegation edges until it either reaches a trusted
principal with compatible scope or exhausts all possible paths.

The inputs to the evaluation procedure are:

\begin{itemize}
  \item the cleaned token set $T_{\text{clean}}$,
  \item the subject token $t \in T_{\text{clean}}$ to be evaluated,
  \item the trusted principal set $R$,
  \item the requested scope $s_{\text{req}} \subseteq S$.
\end{itemize}

The goal is to determine whether $t$ is authorized for all of the capabilities
in $s_{\text{req}}$, relative to the configured trusted principals.

\subsubsection{Trusted principals} 
\label{sec:trusted-principals}
Let $\mathsf{Id}$ be the set of identities and $S$ be the scope space. A
trusted principal is an identity together with the maximum scope that the
verifier is willing to grant that identity. Formally, the verifier provides a
finite set

\begin{equation}
R \subseteq \mathsf{Id} \times S
\end{equation}

where each pair $(i, s\_r) \in R$ means that identity $i$ is trusted, but only
for capabilities contained within $s\_r$. No identity is implicitly trusted for
all scopes. Trust is always bounded by an explicit root scope.

\subsubsection{Delegation paths to the subject} 
Let $G(T_{\text{clean}}) = (V, E)$ be the capability graph defined in Section
\ref{sec:capability-graph-construction}, with $V = \{ u \in T_{\text{clean}} \mid \mathsf{kind}(u) \in \{\text{attest}, \text{vouch}\} \}$.

A \textbf{delegation path for $t$} is a finite
sequence of nodes

\begin{equation}
p = (n_0, n_1, \cdots, n_k)
\end{equation}

such that:

\begin{itemize}
  \item $n_k = t$,
  \item $(n_j, n_{j+1}) \in E$ for all $0 \le j < k$.
\end{itemize}

A path $p$ is rooted in a trusted principal $(i, s_r) \in R$ if the first node
in the path is a token issued by that identity:

\begin{equation}
\mathsf{iss}(n_0) = i.
\end{equation}

Only such paths are relevant for capability evaluation. Paths in the graph that
do not terminate at $t$, or whose initial issuer is not in $R$, are ignored for
the purpose of deciding whether $t$ is authorized.

Let $\mathsf{Vouch}(p)$ denote the set of vouch tokens that appear along the path:

\begin{equation}
\mathsf{Vouch}(p) = \{ n_j \mid 0 \le j < k,\ \mathsf{kind}(n_j) = \text{vouch} \}.
\end{equation}

\subsubsection{Path scope and attenuation} 

Scopes are sets of capabilities, and both delegation statements and terminal
statements may constrain the authority conveyed along a delegation path. Each
vouch token $v$ carries an explicit scope value $\mathsf{scope}(v) \in S$, and
each trusted principal $(i, s_r)$ contributes a root scope $s_r$ that bounds the
maximum authority initially granted by the verifier.

Although attestation tokens do not introduce delegation edges, they may assert
scope constraints that limit the applicability of the statement they represent.
Such constraints participate in scope attenuation at the terminal node of a
delegation path.

Let $\mathsf{tscope}(u) \in S$ denote the scope asserted by a token $u$. For both
vouch and attestation tokens, $\mathsf{tscope}(u) = \mathsf{scope}(u)$. If an
attestation token does not assert an explicit scope constraint, it is treated
as unconstrained with respect to scope, so that $\mathsf{tscope}(u) = S$.   

In practice, capability evaluation begins at the subject token $t$, which is
typically an attestation. If the subject asserts a scope constraint, that
constraint bounds all authority derivable from any delegation path terminating
at $t$. Subsequent vouch tokens along the path may further attenuate authority,
and the root scope $s_r$ provides an overall upper bound.

Given a trusted principal $(i, s_r) \in R$ and a delegation path

\begin{equation}
p = (n_0, n_1, \cdots, n_k = t),
\end{equation}

that is rooted in $(i, s_r)$, the effective scope of the path is defined as:

\begin{equation}
\mathsf{scope}(p) =
s_r
\cap
\bigcap_{v \in \mathsf{Vouch}(p)} \mathsf{scope}(v)
\cap
\mathsf{tscope}(t).
\end{equation}

If $\mathsf{Vouch}(p)$ is empty, the intersection over vouch tokens is taken to be
the entire scope space, so that the effective scope reduces to

\begin{equation}
\mathsf{scope}(p) = s_r \cap \mathsf{tscope}(t).
\end{equation}

A delegation path $p$ is said to be \textbf{admissible} if its effective scope is
non-empty:

\begin{equation}
\mathsf{scope}(p) \neq \varnothing .
\end{equation}

Scope attenuation is monotonic. Extending a path by adding additional vouch
tokens can only reduce or preserve the effective scope; it can never expand it.
Attestation tokens similarly restrict scope without propagating authority. This
ensures that authority cannot grow through delegation or combination of
statements, and that all effective capabilities are bounded by the most
restrictive constraint encountered along the path.

\subsubsection{Acceptance for a requested scope} 
A request is expressed as a set of required capabilities

\begin{equation}
s_{\text{req}} \subseteq S.
\end{equation}

The evaluator must decide whether the subject token $t$ is authorized for at
least the capabilities in $s_{\text{req}}$ under the current trust
configuration and cleaned token set.

A token $t \in T_{\text{clean}}$ is \textbf{accepted} for the requested scope
$s_{\text{req}}$ if and only if there exists at least one delegation path $p$
in $G(T_{\text{clean}})$ such that:

\begin{enumerate}
  \item $p$ terminates at $t$,
  \item $p$ is rooted in some trusted principal $(i, s_r) \in R$,
  \item the path scope satisfies the minimum required scope:
\begin{equation}
s_{\text{req}} \subseteq \mathsf{scope}(p).
\end{equation}
\end{enumerate}

The acceptance condition evaluates each delegation path independently. Scopes
from different paths are never combined or merged. A token is authorized for
the request \textbf{if and only if a single admissible path from a trusted
principal to $t$ preserves all capabilities in $s_{\text{req}}$}. If no such
path exists, the model determines that $t$ is not authorized for the requested
scope.

Define the acceptance predicate:

\begin{equation}
\mathsf{Accept}(T_{\mathrm{clean}}, t, R, s_{\mathrm{req}}) \iff
\exists p \text{ such that }
p \text{ is a path rooted in some } (i, s_r) \in R,
p \text{ terminates at } t,
\text{ and } s_{\mathrm{req}} \subseteq \mathsf{scope}(p).
\end{equation}

\subsubsection{Evaluation function and determinism} 
The capability evaluation function answers the question of whether a specific
token supports a requested scope, given a cleaned token set and a trust
configuration. It is defined as:

\begin{equation}
\mathsf{Eval}(T_{\mathrm{clean}}, t, R, s_{\mathrm{req}}) =
\begin{cases}
\text{accept} & \text{if } \mathsf{Accept}(T_{\mathrm{clean}}, t, R, s_{\mathrm{req}}) \text{ holds}\\
\text{reject} & \text{otherwise}
\end{cases}
\end{equation}

All components involved in $\mathsf{Eval}$ are determined solely by:

\begin{itemize}
  \item $T_{\text{clean}}$, produced by the state resolution rules of Section \ref{sec:state-resolution},
  \item the subject token $t$,
  \item the trusted principal set $R$,
  \item the requested scope $s_{\text{req}}$.
\end{itemize}

Graph construction from $T_{\text{clean}}$ is deterministic, path scopes are
computed by fixed set intersections, and the acceptance predicate is a pure
logical condition over paths. Therefore capability evaluation is deterministic:
If two evaluators receive identical inputs $T_{\text{clean}}, t, R,
s_{\text{req}}$, they construct the same graph, identify the same set of
candidate paths, compute the same path scopes, and reach the same accept or
reject decision.

\subsubsection{Implementation considerations} 
The ZI-CG model does not prescribe a particular search strategy for discovering
paths. However, the attenuating nature of scope offers natural opportunities
for efficient evaluation. During a backward search from $t$, an implementation
may maintain a current accumulated scope along each partial path. As soon as
this accumulated scope becomes empty, or no longer contains the requested
scope $s_{\text{req}}$, that partial path can be abandoned. This early
termination does not alter the semantics of $\mathsf{Eval}$; it simply avoids
exploring paths that cannot possibly satisfy the acceptance condition.

In summary, capability evaluation in a ZI-CG system is a local, goal-directed
computation that starts at the subject token and searches for a single
admissible delegation chain from a scoped trusted principal. Authority never
expands by combining multiple paths. It only attenuates along individual paths
that survive state resolution and remain compatible with the requested scope.

\subsection{Other deterministic evaluators} 
Although capability checking is the primary evaluation described in this model,
ZI-CG does not restrict evaluators to that specific query. Any computation over
the graph $G(T_{\text{clean}})$ is ZI-CG-compliant provided it satisfies
three conditions:

\begin{enumerate}
  \item \textbf{Locality} - results depend only on $T_{\text{clean}}$, the verifier's
trusted principals, and explicit query parameters.

  \item \textbf{Determinism} - identical inputs must yield identical outputs.

  \item \textbf{Scope attenuation} - authority derived from delegation must never expand
beyond what a single admissible path permits.

\end{enumerate}

Within these constraints, evaluators may perform alternative analyses, such as
enumerating valid delegation paths, examining remaining scopes along each path,
or performing reachability or auditing queries. These computations do not alter
authorization semantics; they illustrate that ZI-CG provides a general
substrate for deterministic trust reasoning based solely on the local token
set.

\section{Vouchsafe: A complete instantiation of the ZI-CG model} 
Vouchsafe is a working system that implements the Zero-Infrastructure
Capability Graph (ZI-CG) model in a concrete, deployable form. It uses only
widely available primitives: identities are encoded as URNs, tokens are
serialized as JSON Web Tokens (JWTs) \cite{Jones2015JWT} with JWS signatures
\cite{Jones2015JWS}, content addressing uses SHA-256, and identities and
signatures use Ed25519 \cite{Jones2016EdDSA}. The reference implementation is
written in JavaScript and runs in modern web browsers and Node.js, which
demonstrates that a ZI-CG system can operate on ordinary client devices without
specialized infrastructure.

This section describes how Vouchsafe instantiates the abstract mechanisms of
Section \ref{sec:formal-model}. The goal is not to provide a full format specification, but to show
how each requirement of the model is realized in practice.

\subsection{Identity construction} 
\label{sec:identity-construction}
Vouchsafe identities are encoded as URNs of the form

\begin{equation}
\texttt{urn:vouchsafe:<label>.<hash>}
\end{equation}

where \texttt{<label>} is a human meaningful identifier and \texttt{<hash>} is a collision
resistant digest derived from the issuer's public key. In the implementation,
\texttt{<hash>} is computed as a lowercase, unpadded \texttt{base32} encoding (RFC-4648 \S 6) \cite{Josefsson2006}
of the \texttt{SHA-256} hash of the issuer's \emph{raw public key bytes}.
The raw public key bytes are extracted from the DER-encoded public key and
exclude all encoding metadata. The label is not included in the hash; it is
metadata that does not affect the identity binding.

Every Vouchsafe token includes two claims:

\begin{itemize}
  \item \texttt{iss}: the issuer identity, which must be a Vouchsafe URN of the form \texttt{urn:vouchsafe:<label>.<hash>}.
  \item \texttt{iss\_key}: the issuer's public key, encoded as a DER-encoded \texttt{SubjectPublicKeyInfo} structure and base64 encoded for inclusion in the token.
\end{itemize}

In the notation of Section \ref{sec:identities}, Vouchsafe implements

\begin{equation}
i = \mathsf{id}(pk, m)
\end{equation}

by constructing the URN from the label metadata $m$ and the hash of the public
key $pk$, and implements

\begin{equation}
\mathsf{validate}(i, pk)
\end{equation}

by:

\begin{enumerate}
  \item Parsing \texttt{iss} to recover \texttt{<hash>}.
  \item Decoding \texttt{iss\_key} to obtain the DER-encoded public key.
  \item Extracting the raw public key bytes from the DER structure.
  \item Computing $\text{base32}_{\text{lc,np}}(\mathsf{SHA256}(\text{raw}(pk)))$.
  \item Checking that the computed value matches \texttt{<hash>}.
\end{enumerate}

Hashing only the extracted raw public key bytes ensures that identity
derivation is invariant under semantically equivalent DER encodings, and
depends solely on the cryptographic key material.

During token validation, the library verifies the JWS signature using \texttt{iss\_key}
and then checks that \texttt{iss\_key} satisfies this URN binding. A token is accepted
as self verifying only if its signature is valid and its \texttt{iss} and \texttt{iss\_key}
satisfy the identity binding function. This realizes the self resolving
identity requirement of Section \ref{sec:requirements} and the identity model of Section \ref{sec:identities} in a
fully offline way.

\subsection{Token encoding and signing} 
\label{sec:token-encoding}
Vouchsafe encodes all ZI-CG tokens as signed JWTs using JWS with Ed25519
signatures. The JWT payload (the token body) contains both the issuer identity
and public key, along with the minimal set of fields needed to realize the
ZI-CG token structure.

At a structural level, the body has the form

\begin{equation}
\mathsf{body}(t) = \bigl(\mathsf{iss}(t), \mathsf{pk}(t), \mathsf{payload}(t)\bigr)
\end{equation}

where:

\begin{itemize}
  \item \texttt{iss} encodes $\mathsf{iss}(t)$ and must be a Vouchsafe URN.

  \item \texttt{iss\_key} encodes $\mathsf{pk}(t)$, the public key corresponding to \texttt{iss}.

  \item \texttt{payload} contains:

  \begin{itemize}
    \item \texttt{kind}: a string beginning with \texttt{vch:} that identifies the token type (attestation, vouch, revocation, or burn).

    \item \texttt{jti}: a mandatory, issuer-unique token identifier (UUID). Reuse by the same issuer is forbidden. The pair (\texttt{iss}, \texttt{jti}) identifies a specific statement act and is used as the target of revocation and for application-level reference.

    \item \texttt{sub}: the \texttt{jti} of the subject statement. Its interpretation is fixed by token kind, as specified below.

    \item Type specific claims that realize the fields from Section \ref{sec:token-type-semantics}:

    \begin{itemize}
      \item \texttt{vch\_iss}, \texttt{vch\_sum} for vouch and revocation tokens, which jointly identify a subject token by issuer and content hash.
      \item \texttt{revokes} for revocation tokens, which identifies the prior token being withdrawn.
      \item \texttt{burns} for burn tokens, which identifies the identity being terminated.
      \item \texttt{purpose} for vouch and attestation tokens, which encodes the scope as a set of capability strings.
    \end{itemize}
  \end{itemize}
\end{itemize}

Let $\mathsf{jwt}(t) \in \{0,1\}^\ast$ denote the exact JWS compact serialization
string of token $t$ as transmitted on the wire. The content-addressed token
identifier is computed as

\begin{equation}
\mathsf{tid}(t) = \mathsf{SHA256}\bigl(\mathsf{jwt}(t)\bigr).
\end{equation}

Because $\mathsf{jwt}(t)$ includes the protected header, payload, and signature,
any change to the token contents, signing key, or signature invalidates both
the signature verification and the content identifier.

Subject tokens referenced by vouches and revocations are identified by
(\texttt{sub}, \texttt{vch\_iss}, \texttt{vch\_sum}), where \texttt{vch\_sum}
is the hexadecimal encoded \texttt{SHA-256} hash of the subject token's exact
JWS compact serialization string as originally issued and transmitted, not a
re-encoding or reconstructed representation. This implements the abstract
$\mathsf{ref}(t)$ field from Section \ref{sec:vouch-tokens} and ensures that
subject references are content addressed.

Standard JWT claims such as \texttt{iat}, \texttt{nbf}, or \texttt{exp} may be present and are used
by applications for temporal policy, but they are not required by the ZI-CG
model and do not participate directly in graph construction or capability
evaluation. The core evaluator depends only on the identity fields (\texttt{iss},
\texttt{iss\_key}), the issuer-local statement identifier \texttt{jti}
(together with \texttt{iss}), the type specific claims (\texttt{sub}, \texttt{vch\_iss},
\texttt{vch\_sum}, \texttt{revokes}, \texttt{burns}, \texttt{purpose}), and the
signature.

With this structure, Vouchsafe realizes the self verifying token abstraction of
Section \ref{sec:tokens}: the origin and integrity of each token can be validated using only
its contents and the standard JWS signature semantics, and all references
needed for graph construction and state resolution are encoded as content
addressed fields inside the token body.

\subsubsection{Statement and subject identification invariants}

Every Vouchsafe token contains a mandatory \texttt{jti} field whose value is a
fresh UUID. Reuse of a \texttt{jti} by the same issuer is forbidden. The pair
(\texttt{iss}, \texttt{jti}) uniquely identifies a specific statement issued by
that identity.

The \texttt{sub} field is used uniformly across all token kinds to identify the
\emph{subject statement} by its \texttt{jti}. For attestation and burn tokens,
the statement is self-referential, and \texttt{sub} equals the token's own
\texttt{jti}. For vouch and revocation tokens, \texttt{sub} equals the
\texttt{jti} of the subject token being endorsed or discussed.

Revocation tokens distinguish between the statement being withdrawn and the
subject of that statement. The \texttt{revokes} field identifies the prior
statement act by its \texttt{jti}, while the triple
(\texttt{sub}, \texttt{vch\_iss}, \texttt{vch\_sum}) identifies the subject token
to which that statement applied. A revocation applies only if both the statement
identifier and the subject reference match exactly.

The separation between statement identity (\texttt{iss}, \texttt{jti}) and
content identity (\texttt{sub}, \texttt{vch\_iss}, \texttt{vch\_sum}) allows
revocation of a specific issuer act while simultaneously preventing redirection
to a different subject or to a modified version of the original token.

\subsection{Token types and correspondence with the model} 
All Vouchsafe tokens share the common structure from Section \ref{sec:token-encoding}: \texttt{iss},
\texttt{iss\_key}, \texttt{kind}, \texttt{jti}, \texttt{sub}, and a set of type specific claims. Arbitrary
additional claims may appear for application use and are ignored by the core
evaluator. This section describes only the fields needed to instantiate the
four token categories defined in Section \ref{sec:token-type-semantics}.

\subsubsection{Attestation tokens} 
Attestation tokens carry \texttt{kind = "vch:attest"}. The \texttt{sub} field identifies the
token itself and must equal the \texttt{jti}, and additional payload
claims encode statements about that subject.

In the model's terms, these tokens have $\mathsf{kind}(t) = \text{attest}$.
Their type specific content appears only as data attached to a node in the
graph; the evaluator's task is to decide whether that node is trusted, not to
interpret the attested claims.

\subsubsection{Vouch tokens} 
Vouch tokens implement delegated trust and carry \texttt{kind = "vch:vouch"}. They use three fields that matter for the model:

\begin{itemize}
  \item \texttt{sub}: the \texttt{jti} of the subject token being endorsed,
  \item \texttt{vch\_iss}: the identity of that subject token's issuer,
  \item \texttt{vch\_sum}: the SHA-256 hash of the subject token's exact on-the-wire JWS compact serialization,
\end{itemize}

Together, (\texttt{sub}, \texttt{vch\_iss}, \texttt{vch\_sum}) identify a single subject token in
$T_{\text{valid}}$. This realizes the abstract reference $\mathsf{ref}(t)$
from Section \ref{sec:vouch-tokens} and prevents substitution of a different token with similar
content. The \texttt{purpose} claim provides the delegation scope, interpreted as a
set of capability strings and instantiating $\mathsf{scope}(t) \subseteq S$.

In the model's notation, vouch tokens implement

\begin{equation}
\mathsf{typefields}(t) = \bigl(\mathsf{ref}(t), \mathsf{scope}(t)\bigr)
\end{equation}

with $\mathsf{ref}(t)$ derived from (\texttt{sub}, \texttt{vch\_iss}, \texttt{vch\_sum}) and
$\mathsf{scope}(t)$ derived from \texttt{purpose}. Vouch tokens that survive into
$T_{\text{clean}}$ induce directed edges in the capability graph annotated
with these scopes, as described in Section \ref{sec:capability-graph-construction}.

Because subject references are content-addressed, a vouch token can only
reference tokens that already exist at issuance time, ensuring that the
capability graph is acyclic by construction.

\subsubsection{Revocation tokens} 
Revocation tokens withdraw earlier statements and carry \texttt{kind = "vch:revoke"}. They use:

\begin{itemize}
  \item \texttt{sub}: the \texttt{jti} of the subject token whose endorsements or assertions are being reconsidered,
  \item \texttt{vch\_iss}: the identity of that subject token's issuer,
  \item \texttt{vch\_sum}: the SHA-256 hash of the subject token's exact on-the-wire JWS compact serialization,
  \item \texttt{revokes}: the \texttt{jti} of the specific prior vouch or attestation being revoked.
\end{itemize}

The triple (\texttt{sub}, \texttt{vch\_iss}, \texttt{vch\_sum}) instantiates $\mathsf{ref\_subject}(t)$ in Section \ref{sec:revocation-tokens}, while \texttt{revokes} instantiates $\mathsf{revoke\_target}(t)$. A revocation token matches an earlier statement only if:

\begin{enumerate}
  \item \texttt{revokes} matches the earlier token's \texttt{jti}, and
  \item (\texttt{sub}, \texttt{vch\_iss}, \texttt{vch\_sum}) match the subject token referenced by that earlier statement.
\end{enumerate}

During construction of $T_{\text{clean}}$, any matched vouch or attestation
is removed. This is the concrete implementation of the proof of omission based
revocation semantics from Section \ref{sec:state-resolution}.

\subsubsection{Burn tokens} 
Burn tokens terminate an identity and carry \texttt{kind = "vch:burn"}.
They are strictly self-directed statements and use a single model-relevant field:

\begin{itemize}
  \item \texttt{burns}: the issuer identity \texttt{iss}.
\end{itemize}

A burn token is valid only if \texttt{burns} is exactly equal to the token's
issuer identity \texttt{iss}. Burn tokens that name any other identity are
invalid and must be rejected during structural validation.

As with attestation tokens, burn tokens are self-referential: the \texttt{sub}
field equals the token's own \texttt{jti}. A burn therefore represents an
identity's explicit declaration that it must no longer be trusted.

In the model's notation, burn tokens implement
$\mathsf{burn}(t) = \mathsf{iss}(t)$ as defined in
Section~\ref{sec:burn-tokens}. During state resolution, if a valid burn token is
present in $T_{\text{valid}}$, the evaluator removes all tokens whose issuer
identity $\mathsf{iss}$ matches the burned identity before constructing
$T_{\text{clean}}$.

This constraint ensures that identity termination is strictly self-sovereign.
No identity can revoke, suspend, or terminate any other identity. There is no
mechanism in the ZI-CG model or in Vouchsafe by which authority over identity can
be delegated or exercised by third parties.

\subsubsection{Mapping between ZI-CG model elements and Vouchsafe fields} 
To make the correspondence between the abstract ZI-CG model and the concrete
Vouchsafe implementation explicit, Table~\ref{tab:zicg-vouchsafe-mapping}
summarizes how model-level concepts are realized as JWT claims.

\begin{table}[h]
\centering
\begin{tabular}{|l|l|l|}
\hline
\textbf{ZI-CG concept} & \textbf{Vouchsafe field(s)} & \textbf{Notes} \\
\hline
Identity $i \in I$ &
\texttt{iss} &
Vouchsafe URN derived from issuer public key. \\
\hline
Public key $pk \in PK$ &
\texttt{iss\_key} &
DER-encoded public key, base64 encoded in JWT. \\
\hline
Token $t$ &
JWS compact JWT &
Signed, self-contained statement. \\
\hline
Token identifier $\mathsf{tid}(t)$ &
$\mathsf{SHA256}(\mathsf{jwt}(t))$ &
Hash of exact on-the-wire JWS compact serialization. \\
\hline
Statement identifier &
(\texttt{iss}, \texttt{jti}) &
Issuer-local identifier of a statement act. \\
\hline
Subject reference $\mathsf{ref}(t)$ &
(\texttt{sub}, \texttt{vch\_iss}, \texttt{vch\_sum}) &
Content-addressed reference to a subject token. \\
\hline
Token kind $\mathsf{kind}(t)$ &
\texttt{kind} &
\texttt{vch:attest}, \texttt{vch:vouch}, \texttt{vch:revoke}, \texttt{vch:burn}. \\
\hline
Delegation scope $\mathsf{scope}(t)$ &
\texttt{purpose} &
Set of capability strings. \\
\hline
Revoked statement &
(\texttt{iss}, \texttt{revokes}) &
Identifies prior statement act being withdrawn. \\
\hline
Burned identity &
\texttt{burns} &
Identity being terminated; must equal \texttt{iss}. \\
\hline
\end{tabular}
\caption{Mapping from ZI-CG abstract model elements to Vouchsafe JWT fields.}
\label{tab:zicg-vouchsafe-mapping}
\end{table}

This mapping makes explicit the separation between statement identity and
content identity in Vouchsafe. The \texttt{jti} field identifies a specific
statement act by an issuer, while \texttt{vch\_sum} commits to the exact encoded
content of a referenced token. Together with issuer identity, these fields
instantiate the ZI-CG notions of reference, delegation, revocation, and burn
without reliance on external resolution or mutable infrastructure.

\subsection{State resolution and construction of $T_{\text{clean}}$} 
Vouchsafe implements the ZI-CG state resolution procedure using the same three
stage structure defined in Section \ref{sec:state-resolution}. The evaluator begins with an
application supplied token set $T$, decodes and validates each token, producing
$T_{\text{valid}}$, collects all revokes and burns, and then constructs
$T_{\text{clean}}$ by omitting any token whose validity is contradicted by the
provided statements. This yields the effective state on which graph
construction and capability evaluation operate.

\subsubsection{Validation and construction of $T_{\text{valid}}$} 
Each token in $T$ is first decoded using standard JWT processing. Validation succeeds only if:

\begin{enumerate}
  \item The embedded signature verifies against \texttt{iss\_key}.
  \item The identity binding is correct:

\begin{equation}
\mathsf{iss} = \texttt{urn:vouchsafe:}\mathrm{label}\ .\ 
\mathsf{base32}\bigl(\mathsf{SHA256}(\mathsf{raw}(\texttt{iss\_key}))\bigr)
\end{equation}

where the hash is recomputed from the raw public key bytes extracted from
\texttt{iss\_key} using the same algorithm and encoding defined in Section
\ref{sec:identity-construction}.
\end{enumerate}

Tokens that fail either check are excluded. The remaining tokens form
$T_{\text{valid}}$. 

Only tokens that appear in $T_{\text{valid}}$ participate in revocation and
burn semantics; invalid tokens have no effect on state resolution.

\subsubsection{Extraction of revocations and burns} 
From $T_{\text{valid}}$, the evaluator extracts two classes of control tokens
that influence state resolution:

\begin{itemize}
  \item \textbf{Burn tokens}, which terminate an identity and suppress all
  statements issued by that identity.
  \item \textbf{Revocation tokens}, which retract specific earlier statements
  issued by the same identity.
\end{itemize}

A revocation token is interpreted using two distinct references that must
jointly match a prior statement.

First, the subject of the revoked statement is identified by the triple:

\begin{itemize}
  \item \texttt{sub}: the \texttt{jti} of the subject token,
  \item \texttt{vch\_iss}: the issuer identity of that subject token,
  \item \texttt{vch\_sum}: the SHA-256 hash of the subject token's exact JWS
  compact serialization string as originally issued.
\end{itemize}

Together, these fields identify the exact subject token being referenced,
committing both to its issuer and to its full on-the-wire encoded content. This
prevents a revocation from being redirected to a different subject or to a
modified version of the same token.

Second, the specific prior statement being withdrawn is identified by:

\begin{itemize}
  \item \texttt{revokes}: the \texttt{jti} of the vouch or attestation token
  being revoked.
\end{itemize}

A revocation token matches an earlier statement only if:

\begin{enumerate}
  \item the issuer of the revocation is identical to the issuer of the
  referenced statement, and
  \item the \texttt{revokes} value matches the statement's \texttt{jti}, and
  \item the subject reference (\texttt{sub}, \texttt{vch\_iss},
  \texttt{vch\_sum}) matches exactly the subject token referenced by that
  statement.
\end{enumerate}

Only when all of these conditions hold does the revocation apply. During
construction of $T_{\text{clean}}$, any matched vouch or attestation token is
excluded from the effective state.

Burn tokens are interpreted independently. A burn token specifies an identity
whose entire statement history must be suppressed. When a valid burn token is
present, all tokens issued by the burned identity are removed prior to graph
construction.

Only revocation and burn tokens that appear in $T_{\text{valid}}$ participate
in state resolution. Invalid, malformed, or unauthenticated control tokens have
no effect. This realizes the model's proof-of-omission semantics using only
locally available signed data.

\subsubsection{Construction of $T_{\text{clean}}$} 
A token $t \in T_{\text{valid}}$ is included in the cleaned token set
$T_{\text{clean}}$ if and only if it is neither burned nor revoked under the
state resolution rules.

Formally, a token $t \in T_{\text{valid}}$ is included in $T_{\text{clean}}$ if
and only if all of the following conditions hold:

\begin{enumerate}
  \item \textbf{The token is not issued by a burned identity.}

  If there exists a burn token $b \in T_{\text{valid}}$ such that
  $\texttt{burns}(b) = \texttt{iss}(t)$, then $t$ is excluded from
  $T_{\text{clean}}$.

  \item \textbf{The token is not targeted by an applicable revocation token.}

  A revocation token $r \in T_{\text{valid}}$ applies to a prior token $t$ if and
  only if all of the following conditions hold:

  \begin{itemize}
    \item $t$ is a vouch or attestation token,
    \item $\texttt{iss}(r) = \texttt{iss}(t)$,
    \item \texttt{revokes}$(r) = \texttt{jti}(t)$,
    \item the subject reference matches exactly, meaning that
    \texttt{sub}, \texttt{vch\_iss}, and \texttt{vch\_sum} in $r$ match the
    corresponding subject reference carried by $t$.
  \end{itemize}

  If such a revocation token exists, $t$ is excluded from $T_{\text{clean}}$.
\end{enumerate}

All tokens in $T_{\text{valid}}$ that satisfy both conditions remain in
$T_{\text{clean}}$. Tokens excluded by either burn or revocation are omitted
entirely and do not participate in subsequent graph construction or capability
evaluation.

This procedure implements the ZI-CG proof-of-omission rule: a statement remains
effective precisely when the supplied token set does \emph{not} contain any
contradictory signed statement issued by the same identity. 

This computation-only process results in a $T_{\text{clean}}$ that represents
the complete and self-contained view of the current state of all trust-relevant
statements within the provided token set.

\subsection{Capability graph construction} 
Once $T_{\text{clean}}$ has been computed, the Vouchsafe evaluator constructs
the capability graph $G = (V, E)$ as follows.

\subsubsection{Nodes}
All \textbf{remaining, unrevoked and unburned tokens} are added as nodes:

\begin{equation}
V = \{ u \in T_{\text{clean}} \mid \mathsf{kind}(u) \in \{\text{attest}, \text{vouch}\} \}
\end{equation}

Revocation and burn effects have already been applied during construction of
$T_{\text{clean}}$. Tokens whose statements were withdrawn, or whose issuers
were burned, were removed at that stage and therefore never appear in the
graph. The nodes of $G$ are exactly the statements that remain valid under the
model's state resolution rules.

\subsubsection{Edges}
Edges are induced only by vouch tokens. For each vouch token $t \in
T_{\text{clean}}$, the triple (\texttt{sub}, \texttt{vch\_iss},
\texttt{vch\_sum}) identifies a subject token $u \in T_{\text{clean}}$. If
multiple tokens share the same jti but differ in encoding, only the one whose
encoded JWT hash matches vch\_sum is eligible. If such a token exists, the
evaluator adds a directed edge $ t \rightarrow u $ labeled with the delegation
scope derived from the \texttt{purpose} claim of $t$. If no matching subject
token is present in $T_{\text{clean}}$, the vouch has no effect and contributes
no edge.

\subsubsection{Result}
The resulting directed graph $G(T_{\text{clean}})$ contains exactly the
surviving statements and delegations, with their associated scopes. At this
point, the graph embodies all information required for a correct trust
evaluation under the ZI-CG model; the capability evaluation engine operates
purely by traversing this graph and intersecting scopes, without further
modification of the token set.

The use of content addressed references enforces acyclicity by construction.
For each vouch token, \texttt{vch\_sum} is defined as the SHA-256 hash of the complete
encoded JWT string of the subject token. This requires the subject token to be
fully constructed before the vouch can be created. As a result, edges in
$G(T_{\text{clean}})$ always point from later created tokens toward earlier
tokens whose encodings were already fixed. Under standard assumptions for
SHA-256, this prevents the construction of mutually referential token sets, so
the capability graph $G(T_{\text{clean}})$ is a directed acyclic graph for
any valid Vouchsafe token set.

\subsection{Capability evaluation in Vouchsafe}

Once the capability graph $G(T_{\text{clean}})$ has been constructed, Vouchsafe
performs capability evaluation by traversing the graph and intersecting scope
constraints, exactly as specified by the ZI-CG evaluation model in
Section~\ref{sec:capability-evaluation}. No further modification of the token
set or graph structure occurs during this phase.

In Vouchsafe, scope is represented using the \texttt{purpose} field carried by
attestation and vouch tokens. A purpose value is a whitespace-separated set of
capability labels and is interpreted as a set of scopes. Trusted principals are
configured with a corresponding root purpose set that bounds the maximum
authority initially granted by the verifier.

Evaluation begins at the subject token under consideration and proceeds by
examining delegation paths in $G(T_{\text{clean}})$ that terminate at that
token. Along each such path, the effective purpose is computed by intersecting
the root purpose of the trusted principal, the purpose values asserted by each
vouch token on the path, and any purpose constraint asserted by the terminal
attestation. This corresponds directly to the path scope computation defined in
Section~\ref{sec:capability-evaluation}.

A request is accepted if and only if there exists at least one delegation path
from a trusted principal to the subject token whose effective purpose includes
all requested capabilities. Purpose values from different paths are never
combined. Authorization succeeds only when a single admissible path preserves
the full requested scope.

Because path traversal and purpose intersection operate solely on
$G(T_{\text{clean}})$, the subject token, the configured trusted principals, and
the requested purpose, capability evaluation in Vouchsafe is deterministic.
Any two evaluators that receive identical inputs will compute identical path
scopes and reach the same accept or reject decision.

\section{Security analysis} 
This section analyzes the security properties of the Zero-Infrastructure
Capability Graph (ZI-CG) model and its Vouchsafe instantiation. We state the
security contract, threat model, and residual risks, with guarantees understood
relative to a verifier's local computation over signed tokens under standard
assumptions for Ed25519 and SHA-256.

\subsection{Security contract and scope} 
ZI-CG is designed so that the integrity and provenance of trust statements are
enforced by token structure and local verification, rather than by trusted
services or distribution channels. This subsection summarizes the security
contract and the intended scope of the model.

\textbf{Model-level guarantees.} Under standard assumptions for Ed25519 and
SHA-256, a verifier that correctly implements structural validation and
$T_{\text{clean}}$ construction obtains:

\begin{itemize}
  \item \textbf{Integrity of statements:} tokens cannot be modified without
  invalidating their signatures.
  \item \textbf{Authentic provenance:} each accepted token is bound to an issuer
  identity and public key, preventing key substitution for a fixed identity.
  \item \textbf{Graph immutability:} vouch and revoke tokens bind to their
  subjects by content, preventing retroactive rewriting or splicing of existing
  edges without key compromise or hash collisions.
  \item \textbf{Locality of authority:} tokens confer no ambient authority; a
  statement has effect only relative to a verifier's trusted roots and requested
  scope.
  \item \textbf{Deterministic evaluation:} for a fixed cleaned token set and
  verifier configuration, any correct implementation derives the same trust
  decision, enabling reproducible audits.
\end{itemize}

\textbf{Residual risks and non-goals.} ZI-CG does not attempt to guarantee
semantic truth of application claims, protection against endpoint compromise,
or global agreement on token availability.  Adversaries who control
infrastructure can influence \emph{which valid tokens are available} at
evaluation time, but cannot change what any already issued token says without
key compromise.  As a result, the primary residual attack surface lies in key
compromise, distribution of burn and revocation tokens, and verifier policy
configuration.

\subsection{Threat model} 
\label{sec:threat-model}
We assume an adversary who can:

\begin{itemize}
  \item Observe, inject, modify, and replay messages via any mechanism used to distribute tokens.
  \item Attempt to misrepresent identity, forge statements, or manipulate delegation structure.
  \item Present arbitrary token sets to a verifier, including large or carefully chosen sets.
  \item Compromise private keys of some principals.
\end{itemize}

We assume no trusted infrastructure and no trusted distribution channels. The
verifier's evaluation logic and local configuration are assumed to be
uncompromised. All other aspects of the environment, including transport,
storage, and token repositories, may be under adversarial control.

\subsubsection{Assumptions and limitations}
This analysis relies on the following explicit assumptions:

\begin{itemize}
  \item \textbf{Cryptographic assumptions.} The Ed25519 signature scheme is
  unforgeable under chosen message attacks, and SHA-256 is collision resistant
  and preimage resistant.

  \item \textbf{Correct evaluation.} The verifier correctly implements token
  structural validation and the construction of the cleaned token set
  $T_{\text{clean}}$, including burn and revocation processing. Omitting these
  rules changes the model and can invalidate the guarantees stated in this
  section.

  \item \textbf{Token availability.} The model does not specify how tokens are
  distributed or synchronized. An adversary who can suppress or delay delivery
  of relevant burn or revocation tokens can cause verifiers to continue
  accepting statements that would be rejected under a more complete token set.

  \item \textbf{Policy configuration.} Trusted principal sets and initial scopes
  are local policy. If these are misconfigured, the evaluation procedure will
  faithfully compute incorrect authorization outcomes with respect to the
  verifier's intended policy.
\end{itemize}

Given this threat model, we analyze statement integrity, graph manipulation,
key compromise semantics, and the residual attack surface introduced by token
availability and distribution.

\subsection{Integrity and authenticity of statements}
In ZI-CG, trust-relevant statements are represented as signed tokens. A token is
accepted as structurally valid only if:

\begin{enumerate}
  \item The signature verifies under the embedded issuer public key
        \texttt{iss\_key}, and
  \item The issuer identity \texttt{iss} is correctly bound to \texttt{iss\_key}
        under the Vouchsafe URN derivation rule (Section~\ref{sec:identity-construction}).
\end{enumerate}

Under the cryptographic assumptions stated in Section \ref{sec:threat-model}, an
adversary who does not possess the issuer's private key cannot produce a new
token that validates as being issued by that identity, and cannot modify any
field of an already issued token without invalidating its signature.

These properties guarantee authenticity and provenance of statements, not their
semantic correctness. An issuer may assert misleading, incorrect, or deceptive
application-level claims while still producing structurally valid tokens.
Determining the truth or appropriateness of application-specific claims is
outside the ZI-CG model and must be enforced by application policy or external
verification mechanisms.

\subsection{Graph level attacks} 
\label{sec:graph-level-attacks}

\subsubsection{Substitution and graph rewriting} 
Given structurally valid tokens with authenticated issuers, a natural next attack
is to attempt to splice malicious tokens into an existing delegation graph, for
example by redirecting a vouch to a different subject or by retroactively
changing a subject token's content.

In Vouchsafe, vouch and revocation tokens identify their subject by three fields:

\begin{itemize}
  \item \texttt{sub}: the subject token's \texttt{jti},
  \item \texttt{vch\_iss}: the subject token's issuer identity,
  \item \texttt{vch\_sum}: the SHA-256 hash of the subject token's exact JWS
  compact serialization string as originally issued.
\end{itemize}

To change which token a vouch or revocation applies to, an attacker would need
to produce a different token whose JWT encoding has exactly the same SHA-256
hash and is issued by the same identity. Under the collision resistance
assumption for SHA-256, this is infeasible. Likewise, modifying the subject
token itself would change its encoded representation and therefore \texttt{vch\_sum},
breaking the link.

As a result, once tokens have been issued, the structure of the capability
graph is fixed by their signed contents. An attacker can add new tokens, but
cannot rewrite the meaning of existing ones without either:

\begin{itemize}
  \item obtaining the issuer's private key, or
  \item finding collisions in SHA-256.
\end{itemize}

Further, even if an attacker was able to obtain the issuer's private key, creating a 
new token with any changed content would invalidate \texttt{vch\_sum} of any vouches 
targeting that token, breaking the link and rendering them inert.

\subsubsection{Acyclicity and prevention of recursive delegation}

In ZI-CG and therefore Vouchsafe, all delegation edges are induced by vouch
tokens that refer to their subject via a content-addressed hash of the subject
token's complete encoded JWT string. Because this hash can only be computed
once the subject token is fully constructed, a vouch token can only reference
tokens that logically and temporally precede it. In particular, a token cannot
content-address itself (or a token that depends on it) without requiring a hash
collision in \texttt{vch\_sum}.

As a result, it is impossible to construct mutually referential or cyclic
delegation structures without either breaking the collision resistance of
SHA-256 or precomputing the encoding of a token before it exists. Under the
standard cryptographic assumptions, the resulting capability graph is therefore
a directed acyclic graph.

This property prevents recursive self-endorsement, infinite delegation loops,
and scope amplification through cycles. It is enforced by cryptographic data
dependencies rather than by procedural checks or cycle detection during
evaluation.

\subsubsection{Replay and stale tokens}
Because tokens are self contained and verifiable, an attacker can replay old
tokens as long as those tokens remain structurally valid and have not been
revoked or superseded by application policy. The ZI-CG model deliberately
treats revocation and identity termination as explicit data additions, rather
than implicit rules tied to network state or time.

This yields two properties:

\begin{itemize}
  \item For any fixed token set, evaluation is deterministic. Replaying the same
tokens to different verifiers produces the same result.

  \item Protection against replay of stale authority is provided by burn and
revocation tokens, and by application level constraints such as short
lifetimes or issuance policies, not by infrastructure checks.

\end{itemize}

Tokens in ZI-CG do not carry ambient or global authority. A token's effect is
always evaluated relative to a verifier's locally configured trusted roots and
the requested scope. Replaying a valid token in a different application or
context cannot confer additional authority unless that verifier explicitly
trusts the issuer for the relevant scope.

The model guarantees correctness relative to the token set provided. It is the
responsibility of the surrounding system to ensure that relevant revocations
and burns are made available to verifiers that need them, or that tokens are
issued with bounded lifetimes.

\subsubsection{Token set flooding}
An attacker can attempt a denial of service attack by presenting a very large
token set designed to maximize the work of validation and graph traversal. The
ZI-CG model does not preclude this; any system that accepts unbounded input can
be overloaded.

The Vouchsafe evaluator provides two practical mitigations that do not change
the model:

\begin{itemize}
  \item \textbf{Early rejection based on trust roots.} If the token set contains no tokens
issued by any identity in the verifier's trusted set, the evaluator can fail
early without constructing the full graph.

  \item \textbf{Scoped search based on requested capability.} Evaluation starts from the
target token and requested scope, and stops along any path as soon as the
accumulated scope becomes empty or disjoint. Tokens and paths that cannot
contribute to a successful result are never fully explored.

\end{itemize}

In practice, many deployments will prefer delegation structures that remain
small enough to evaluate efficiently. Independently of typical usage patterns,
implementations may impose conservative limits on token set size, indexing
effort, or graph exploration depth to bound worst-case resource consumption in
adversarial settings. Such limits are operational defenses that bound the workload 
presented to the ZI-CG evaluator.

Beyond sheer token count, an adversary may attempt to construct pathological
delegation structures that maximize traversal depth or delay scope elimination,
such as long, narrowly attenuated delegation chains. These cases represent
worst-case inputs for any graph-based authorization system. ZI-CG does not
eliminate such cases at the model level, but its strictly attenuating scope
semantics enable evaluators to prune infeasible paths early without altering
correctness.

These strategies reduce the cost of evaluating adversarial token sets, but
denial of service remains a practical consideration for implementations.

\subsection{Key compromise, and identity burn} 

If an attacker obtains the private key corresponding to an identity
\texttt{iss}, they can issue arbitrary new tokens that appear to originate from
that identity. As in all signature-based systems, this represents a total
compromise: for any evaluator that has not yet received contradictory
statements, the attacker and the legitimate identity become indistinguishable.

Compromise of a \emph{trusted root} identity is qualitatively more severe than
compromise of a downstream identity. If an attacker obtains the private key of
an identity that appears in the verifier's trusted principal set $R$, the
attacker can issue new attestations and vouches that introduce entirely new
delegation paths rooted at that trusted identity. These paths may authorize
third-party statements or capabilities that the legitimate issuer would never
have endorsed.

ZI-CG provides a mechanism to help mitigate key compromise. An identity whose
key has been compromised may issue a \texttt{burn} token. A \texttt{vch:burn}
token naming \texttt{iss} terminates an identity entirely. When present, all
statements issued by that identity are removed from $T_{\text{clean}}$,
preventing further use of both past and future tokens. Burn is the only
operation that completely prevents a compromised key from continuing to
influence evaluation, including the ability to introduce new delegation paths
from a trusted root. Burn is intentionally irreversible: once a burn token is
observed, no subsequent statement can restore trust in that identity.

Because burn tokens are themselves ordinary signed statements, an attacker who
has obtained the private key can also issue a valid burn token. This creates an
interesting dual effect:

\begin{itemize}
  \item \emph{Denial of service:} the legitimate owner loses the ability to use
        the identity, and no recovery is possible;
  \item \emph{Self-limiting effect:} the attacker simultaneously destroys their
        own ability to continue abusing the compromised identity, since all
        future statements will be excluded by evaluators that observe the burn.
\end{itemize}

This duality follows directly from ZI-CG's semantics: once the private key is
compromised, the model cannot distinguish between legitimate and adversarial
issuance. Identity destruction is therefore both the attack and the mitigation.
No revocation strategy can restore trust in the identity, and no mechanism in
ZI-CG permits "reclaiming" it.

There is intentionally no mechanism by which one identity can terminate
another; identity destruction is a strictly self-issued act.

The practical effectiveness of a burn token depends entirely on dissemination.
Evaluators that do not receive it will continue to treat the compromised
identity's prior statements as valid. This reflects an inherent limitation of
any system that bases trust solely on locally available signed data.

Deployments may choose to constrain the use of powerful identities, shorten
token lifetimes, or propagate burn tokens aggressively to reduce the window of
vulnerability. However, the underlying principle remains: \emph{key compromise
irreversibly destroys the identity, and only a burn token, whether issued by the
legitimate party or by the attacker, can terminate its authority.}

\subsection{Determinism and consistency} 
Because all trust-relevant state is encoded in tokens, and because graph
construction and capability evaluation are pure functions of:

\begin{itemize}
  \item the cleaned token set $T_{\text{clean}}$,
  \item the verifier's trusted principal set $R$, and
  \item the requested scope $s_{\text{req}}$,
\end{itemize}

any two correct implementations will derive the same result for the same
inputs. There is no hidden dependency on network state, synchronized clocks, or
external registries.

This property has two security consequences:

\begin{itemize}
  \item \textbf{Consistency across partitions.} Different verifiers that see the same
token set will reach the same trust decision, even if they are operating in
disconnected or adversarial environments.

  \item \textbf{Auditability.} A trust decision can be reconstructed and audited by 
re-running the evaluation on the recorded token set and configuration. The
absence of external dependencies makes such audits reproducible.

\end{itemize}

Determinism does not prevent an adversary from withholding tokens, but it does
ensure that any disagreement between verifiers is explained by differences in
the data they have seen, not by ambiguity in the model.

\subsubsection{Trust root misconfiguration}

The selection of trusted principals and their associated root scopes is a local
policy decision made by each verifier. If a verifier assigns an overly broad
root scope $s_r$ to a trusted identity, the ZI-CG evaluation will faithfully
propagate authority within $s_r$ through valid delegation paths, potentially
granting unintended capabilities.

As defined in Section \ref{sec:trusted-principals}, trust roots are always
configured as pairs $(i, s_r) \in R$, and $s_r$ explicitly bounds the
capabilities for which identity $i$ is trusted. No identity is implicitly
trusted for all scopes, and ZI-CG therefore does not admit unbounded trust roots.
All authorization results are bounded by the verifier's chosen root scopes.

ZI-CG mitigates the impact of such misconfiguration by enforcing scope
attenuation along each delegation path and by making all trust relationships
explicit and auditable in the token set.  However, it cannot prevent a verifier
from choosing a root scope that is broader than intended.  Incorrect trust root
configuration results in incorrect authorization decisions by design, not by
model failure.

This risk is addressed operationally by careful scope design for trusted roots
and by auditing authorization decisions against the token set that justified them.

\subsection{Token distribution and the residual attack surface} 
\label{sec:token-distribution-attack}
The ZI-CG model deliberately separates trust evaluation from data availability.
All correctness properties in this paper are stated relative to the token set
$T$ that a verifier actually holds. This removes online dependencies from the
evaluation loop, but it does not eliminate the problem of distributing tokens.
Instead, token distribution becomes the primary remaining attack surface.

\subsubsection{Attacks on distribution} 
An adversary who controls transport or storage channels for tokens can attempt
several classes of attack:

\begin{itemize}
  \item \textbf{Suppression:} prevent delivery of some tokens, in particular burn or revocation tokens that would reduce or remove previously granted capabilities.
  \item \textbf{Partitioning:} ensure that different verifiers see different subsets of the global token set, leading to divergent but locally consistent trust decisions.
  \item \textbf{Delay:} slow down dissemination of new statements, so that some verifiers continue to operate on outdated information.
  \item \textbf{Flooding:} present very large or carefully constructed token sets in an attempt to exhaust verifier resources, as discussed in Section \ref{sec:graph-level-attacks}.
\end{itemize}

None of these attacks can cause a verifier to accept a token whose signature or
identity binding is invalid, and none can change the meaning of tokens already
in the verifier's possession. They operate exclusively by influencing \textbf{which
valid tokens are available} when evaluation occurs.

The model treats the mechanisms used to distribute tokens as out of scope. 
Deployments may use any transport or storage mechanism that provides
the level of availability and timeliness the application requires. 
Distribution does not affect token validity, but it can determine how long a
verifier operates with a potentially incomplete token set.

\subsubsection{Effect of the actor-centric presentation model} 
Although distribution remains an attack surface, the actor-centric structure of
ZI-CG significantly constrains what successful attacks can achieve.

In typical deployments, the entity seeking authorization presents the tokens
that justify the requested capability. The evaluator starts from this
actor-provided set and optionally augments it with locally held tokens (for
example, revocations or burns). This has three important consequences.

First, \textbf{withholding tokens cannot increase authority}. All capabilities must
be supported by explicit delegation chains in $T_{\text{clean}}$, and
delegation is attenuating. Omitting a token can only remove possible evidence
of authorization, not create new authorization. There is no way to construct
additional privileges by hiding information.

Second, the primary distribution attack with material effect is \textbf{suppression
or delay of revocation and burn tokens}. If a revocation or burn token is
never delivered to a verifier, that verifier may continue to accept an earlier
delegation that the issuer intended to withdraw. Even in this case, the attack
is bounded:

\begin{itemize}
  \item It cannot extend authority beyond what was originally granted in the existing vouches.
  \item It cannot expand scopes along any path; it can only prevent those scopes from being reduced sooner.
  \item Once a burn or revocation token is present in the verifier's token set, its effect follows deterministically from the evaluation rules and does not depend on any further infrastructure.
\end{itemize}

Third, the actor is naturally incentivized to supply the enabling tokens that
support the desired action. Failure to provide a complete chain simply results
in denial. There is no mechanism by which an actor can gain additional
capability by omitting their own tokens, since any missing link in a chain
makes that chain unusable.

In many deployments, verifiers will maintain their own repositories of
disabling tokens (burns and revocations). These can be obtained through any
distribution mechanism that is convenient for the application, and they only
ever remove authority relative to the actor-provided set. This hybrid approach
preserves the zero-infrastructure evaluation semantics while further limiting
the effect of revocation suppression by the actor or by intermediaries under
their control.

\subsubsection{Summary of residual risk} 
Taken together, these properties mean that:

\begin{itemize}
  \item The remaining attack surface lies in the \textbf{completeness and timeliness} of
token distribution, not in the correctness of the evaluation procedure.

  \item Successful distribution attacks can \textbf{delay or prevent reduction of existing
authority}, but cannot create new capabilities or expand scopes beyond what
was originally delegated and left unrevoked.

  \item Differences in trust decisions between verifiers can be attributed to
differences in the token sets they have received, not to ambiguity in the
model or dependence on mutable infrastructure state.

\end{itemize}

ZI-CG mitigates a broad class of attacks that rely on mutable state or
infrastructure-mediated rewriting of trust by committing trust decisions to
signed, content-bound tokens and deterministic local evaluation. The residual
risk is concentrated in key compromise and token availability. While revocation
and burn provide a way to withdraw authority and terminate compromised
identities, these risks remain under adversarial key theft and under
suppression, delay, or partitioning that prevents timely delivery of revocation
and burn tokens to the verifiers that must apply them.

\section{Potential Applications} 

The ZI-CG model provides offline-verifiable identity, delegation, and
revocation using only the token set available to the evaluator. This section
highlights three representative classes of systems in which ZI-CG eliminates
structural failure modes present in infrastructure-dependent trust models,
materially changing what is feasible or reliable. 

\subsection{Disconnected and Adversarial Environments} 

ZI-CG enables trust evaluation without network access, synchronized clocks, or
contact with external authorities. All information required for identity and
authorization is carried by the presented tokens and evaluated locally. This
supports operation in environments where connectivity is intermittent,
unreliable, or intentionally unavailable.

Such conditions arise in disaster response, temporary field operations,
air-gapped facilities, and environments subject to censorship or infrastructure
interference. In these settings, conventional authorization systems can fail
because trust decisions depend on reaching infrastructure that may be
degraded, untrusted, or unreachable.

These same conditions frequently require coordination across administrative or
organizational boundaries, often without prior federation or shared trust
infrastructure. ZI-CG supports such coordination while preserving local control
over trust decisions.

By committing trust decisions to signed, content-bound tokens and deterministic
local evaluation, ZI-CG allows access control, delegation, and revocation to
continue functioning even when communication is disrupted. 

\subsection{Self-Verifying Statements Without Trusted Channels} 

Attestations in a ZI-CG system contain all information required to authenticate
their origin and integrity, allowing any verifier to validate a statement
offline and without third-party confirmation. Such attestations may be
distributed via any mechanism including untrusted networks, public channels,
physical media, or printed representations such as QR codes.

These properties support the communication of verifiably authentic information
in offline or untrusted environments. Representative uses include associating
authentic statements with digital or physical assets, and inventory or
supply-chain workflows where online verification is unavailable or undesirable.
In these settings, statements may be recorded locally, audited offline, and
synchronized later without weakening the trust guarantees provided by the
original attestation.

\subsection{Decentralized Delegation Without Central Authority} 

ZI-CG supports scoped, attenuated delegation among parties without requiring a
central authority, global namespace, or shared trust infrastructure. Authority
is expressed through explicit, signed statements that may be independently
verified by any evaluator possessing the relevant token set.

This delegation model is well-suited to peer-to-peer and decentralized
applications, where participants must grant and exercise limited authority
without relying on a centralized access-control service. In a ZI-CG system, an
identity may delegate authority to another by issuing a vouch token, which may
itself be further delegated subject to attenuation and revocation rules. Each
participant evaluates effective authority locally, without contacting the
delegator or relying on a globally trusted mint or resolver.

Trust relationships can be established, constrained, or withdrawn dynamically
using portable tokens, allowing decentralized systems to adapt to changing
participants and roles without introducing centralized infrastructure into the
trust path.

\section{Limitations and Directions for Future Work} 

The Zero-Infrastructure Capability Graph (ZI-CG) model is intentionally narrow.
It focuses on self-verifying identities, verifiable statements, capability
delegation and attenuation, revocation, and deterministic evaluation over a set
of tokens. This section describes the resulting limitations and outlines areas
where additional mechanisms or complementary layers may be explored. These
limitations do not undermine correctness of evaluation within the model but
instead delineate the boundaries within which ZI-CG should be interpreted and
deployed.

\subsection{Absence of a Prescribed Distribution Model} 

ZI-CG does not prescribe any mechanism for discovering, distributing, or
storing capability tokens. This is a real limitation: applications must decide
how tokens are conveyed to evaluators. It is, however, also an intentional
aspect of the model. By defining only the trust-evaluation semantics, ZI-CG
keeps distribution concerns separate from authorization correctness.

This separation has practical and research consequences. In deployment, it
allows systems to choose distribution methods appropriate to their environment,
from simple file transfer to more complex synchronization schemes. For
research, ZI-CG provides a stable evaluation substrate that can be paired with
experimental distribution and synchronization models. This allows evaluation to
be fixed, so that trust behavior variations can be mapped directly to changing
distribution parameters. The availability of a functional JavaScript
implementation lowers the barrier to such experimentation, enabling these
studies to be conducted with no more than a modern web browser.

\subsection{Absence of global, up-to-the-second correctness} 

The ZI-CG model guarantees deterministic evaluation with respect to a given
token set. It does not attempt to provide a globally synchronized view of the
world, nor does it encode any notion of a single, up-to-date global truth about
authorization or revocation.

Systems that require stronger freshness or global consistency guarantees must
pair ZI-CG with mechanisms that obtain, maintain, and supply an appropriate
token set to each evaluator.

Further exploration of synchronization mechanisms may reveal patterns that
offer stronger consistency guarantees while remaining compatible with ZI-CG's
offline-first evaluation model.

\subsection{Lifetime management without temporal semantics} 

ZI-CG intentionally excludes time from the model. It does not reference clocks,
timestamps, expiration windows, or freshness guarantees. This preserves
deterministic evaluation in environments where timekeeping infrastructure may
be unavailable, unreliable, or contested.

When a reliable local notion of time exists, applications may layer temporal
constraints on top of ZI-CG as an external policy. However, ZI-CG itself defines
capability semantics without embedding lifetime semantics, avoiding dependence
on synchronized clocks or global time agreement.

Questions of trust duration and expiration in zero-infrastructure environments
remain an application-level concern. Non-temporal or coarsely temporal
approaches to limiting authority remain an open area for further study.

\subsection{Identity versus principal: binding and continuity} 

ZI-CG identities are cryptographically self-verifying actors. The model
describes how such an identity can make statements, delegate, attenuate, or
revoke capabilities, and how evaluators verify those assertions. It does not
address how identities correspond to real-world principals, nor does it define
key rotation, succession, or continuity mechanisms.

This separation is deliberate. On one hand, it enables a minimal and portable
form of principal binding: any principal that controls the corresponding private
key can assert ownership of the identity wherever a signed statement can be
presented. On the other hand, applications that require stronger guarantees
around principal continuity, recovery, succession, or rotation must define
those mechanisms outside the ZI-CG model.

For applications that require it, the question of how best to bind long-lived
principals to cryptographic identities while preserving ZI-CG's
zero-infrastructure assumptions remains an open area for further research.

\subsection{Expressiveness of scope definitions} 

ZI-CG adopts a deliberately simple model of capability scope: each capability is
represented as a text string, and attenuation is performed through set
intersection along delegation chains. This design keeps evaluation fully
deterministic and ensures that authority can only decrease as vouches compose.

While this simplicity provides clarity and strong safety properties, it also
limits the expressiveness of capability definitions. Many real-world
authorization domains involve structured or parameterized capabilities, numeric
constraints, or context-sensitive permissions that do not map cleanly onto
flat strings. 

While many such constraints can be applied at the application
layer, there is an opportunity to explore richer scope representations that
retain the core invariants of ZI-CG, determinism, monotonic attenuation, and
infrastructure-free verification, while allowing more complex descriptions of
authority. Future research may explore which forms of additional expressiveness
can be introduced without compromising these guarantees.

\subsection{Human-facing identifiers and disambiguation} 

ZI-CG does not attempt to define a global, human-meaningful namespace. This is a
deliberate omission: the model concerns itself with self-verifying identities and
their cryptographic semantics, not with how those identities are named,
displayed, or interpreted by humans. Global uniqueness is provided by the full
identity string; human-facing representation is left to applications.

As key-based identifiers become more common, practical systems must address
the problem of human disambiguation. Raw cryptographic identifiers are not
well-suited to direct human comparison, and the model does not prescribe how
identities should be labeled, annotated, or distinguished in user interfaces.

Approaches to human-facing disambiguation, such as naming schemes,
contextual labels, or visual representations derived from an identifier, are
compatible with ZI-CG but remain outside its scope. Understanding how such
mechanisms can aid usability without undermining the security or trust semantics
of self-verifying identity is an open area for further study.

\subsection{Integration with infrastructure-rich systems} 

The ``zero-infrastructure'' requirement in ZI-CG applies specifically to the
evaluator: all information needed for a trust decision must be present within
the provided token set. While ZI-CG enables offline trust systems, the model
does not constrain the broader system in which evaluators operate. ZI-CG-style
semantics are compatible with online deployments and may coexist with
directories, distribution services, or policy engines without altering the core
evaluation guarantees.

This suggests a complementary direction for future study: how ZI-CG trust
semantics can be incorporated into conventional online systems. Self-verifying,
composable tokens may reduce reliance on authoritative intermediaries, improve
auditability, or simplify distributed authorization logic. Conversely, online
infrastructure may provide efficient dissemination of ZI-CG tokens while
preserving offline verifiability at the point of evaluation.

\noindent\textbf{Summary}

The limitations described above follow directly from the design goal of ZI-CG:
to define a complete, offline-first model of identity and trust based entirely
on self-verifying statements. Most of the areas identified represent
intentional boundaries rather than missing functionality. By constraining the
model to locally verifiable statements and deterministic evaluation, ZI-CG
remains simple, analyzable, and robust under constrained or adversarial
conditions.

Future work naturally centers on exploring how best to support and extend this
offline-first trust substrate, including distribution and synchronization
models, alternative approaches to lifetime and expiration, more expressive yet
deterministic scope semantics, and human-facing techniques for disambiguation in
key-derived identity systems. ZI-CG decouples trust evaluation from these
concerns, so these mechanisms can be studied and evolved independently without
entangling or redefining the core trust model.

\section{Conclusion} 

Most modern identity and authorization systems assume online
infrastructure: resolvers, directories, revocation services, synchronized
clocks. These assumptions fail precisely in the environments where secure
coordination is most needed, including damaged networks, compromised
infrastructure, and disconnected field operations. This paper has shown that
identity and trust need not depend on such services. By treating identities as
self-verifying public-key-bound identifiers and modeling authorization as a
graph of signed capability statements, the Zero-Infrastructure Capability Graph
(ZI-CG) model makes trust decisions a property of data rather than of
infrastructure.

We formalized ZI-CG as a capability system over immutable, content-addressed
tokens with monotonic attenuation, explicit revocation, and deterministic
evaluation over a supplied token set. We then instantiated this model in the
Vouchsafe implementation, demonstrating that the necessary primitives can be
realized with contemporary cryptography and deployed in commonplace environments
such as web browsers and commodity compute capable devices. 

The security analysis showed that, under standard assumptions, ZI-CG resists a
range of structural attacks that target conventional infrastructures, including
resolver compromise, mutable directory state, and silent credential rewriting,
while making the remaining attack surface explicit and auditable.

Conceptually, ZI-CG and its Vouchsafe instantiation challenge the assumption
that online infrastructure is a necessary root of trust. The central question
is whether the entire verifiable state of an authorization relationship,
including issuance and revocation, can be encoded into signed tokens so that
evaluators need only depend on the presented token set and local policy and not 
on external services. In this model, evaluators need only a token set and a
local policy configuration to reach the same decision anywhere in the network,
or off the network entirely. Together, the model and implementation demonstrate
that a complete, infrastructure-independent trust substrate is both
theoretically sound and practically realizable.

Many opportunities remain for future work, including exploring richer
capability vocabularies, lifetime semantics compatible with
zero-infrastructure operation, and specialized distribution mechanisms for
large-scale deployments. These are natural areas of extension that build upon
the core ZI-CG model. ZI-CG, and its Vouchsafe instantiation, provide a
complete, self-contained evaluation engine that applications can adopt today.
We hope this work provides a foundation for further research and for building
systems whose trust properties remain intact even when infrastructure does
not.

\bigbreak 

\section{Artifacts}

The ZI-CG model presented in this paper is fully instantiated in the open-source
Vouchsafe specification and reference implementation.

\begin{itemize}
  \item \textbf{Vouchsafe Specification (v2.0.2):} \cite{jay_kuri_2025_18012080}\\
  \url{https://doi.org/10.5281/zenodo.18012080}

  \item \textbf{Vouchsafe JavaScript Implementation (v2.0.2):} \cite{jay_kuri_2025_18012093}\\
  \url{https://doi.org/10.5281/zenodo.18012093}
\end{itemize}

Both artifacts are released under the BSD 3-clause license. The reference
implementation is published on \texttt{npm} and is suitable for use in both
browser and server environments.

\section{Acknowledgements}

I thank Tom Vilot for good-natured but persistent skepticism and candid,
intellectually honest discussions, especially during the early stages of this
work, which led to clearer articulation and strengthened the precision of the
definitions and explanations that followed.

I am also deeply grateful to my wife, Rebecca Kuri, whose patience,
encouragement, and unwavering support made this work possible. Her willingness
to listen through its many iterations and her insistence on balance when it
was needed were essential to sustaining the long effort required.

\bibliographystyle{unsrt}
\bibliography{references}

\end{document}